 \documentclass[final,5pt,times]{elsarticle}

\usepackage{amssymb}
\usepackage{amsmath}

\usepackage{xcolor}
\usepackage{graphicx}
\usepackage{dcolumn}
\usepackage{bm}%
\usepackage[caption=false]{subfig}
\usepackage{fixmath}
\usepackage{epstopdf}

\newcommand{\bra}[1]{\left\langle{#1}\right|}
\newcommand{\ket}[1]{\left|{#1}\right\rangle}

\journal{Annals of Physics}

\begin{document}

\begin{frontmatter}



\title{Comparing the extrinsic orbital Hall effect in centrosymmetric and noncentrosymmetric systems: Insights from bilayer transition metal dichalcogenides}


\author[]{Azadeh Faridi$^{a,*}$}
\author[]{Reza Asgari$^{a,b,\dagger}$}

\affiliation[a]{organization={School of Quantum Physics and Matter, Institute for Research in Fundamental Sciences (IPM)},
            city={Tehran},
            postcode={19538-3351}, 
            country={Iran}}
\affiliation[b]{organization={Department of Physics, Zhejiang Normal University},
            city={Jinhua},
            postcode={321004}, 
            country={China}}

\begin{abstract}
We investigate both intrinsic and extrinsic orbital Hall effects (OHE) in bilayer transition metal dichalcogenides (TMDs) in the presence of short-range disorder using quantum kinetic theory. Bilayer TMDs provide an ideal platform to study the effects of inversion symmetry breaking on transport properties due to their unique structural and electronic characteristics. While bilayer TMDs are naturally inversion symmetric, applying a finite gate voltage to create a bias between the layers effectively breaks this symmetry. Our findings reveal that slightly away from the band edges, the extrinsic OHE eventually becomes the dominant contribution in both inversion-symmetric and asymmetric cases, with its prominence increasing significantly as a function of Fermi energy. Furthermore, we demonstrate that breaking inversion symmetry greatly enhances the extrinsic OHE. This enhancement arises from the fundamentally distinct behavior of orbital angular momentum (OAM) in centrosymmetric systems, where intraband components vanish due to symmetry constraints. As a result, in centrosymmetric systems, only the off-diagonal components of the density matrix contribute to the extrinsic OHE. In contrast, in noncentrosymmetric systems, both diagonal and off-diagonal components play a role. Our study suggests that in experimentally relevant highly doped systems, the OHE becomes predominantly extrinsic in both centrosymmetric and noncentrosymmetric although the contribution is much more pronounced in the latter. Importantly, we infer that even a weakly breaking of inversion symmetry can lead to a dramatic enhancement of the OHE, a finding with significant implications for experimental investigations.
\end{abstract}




\end{frontmatter}




\section{Introduction}
\label{introduction}

The pursuit of customized microelectronic devices for information processing has opened new avenues exploring novel degrees of freedom in solids beyond the charge of electrons. Among these, the spin angular momentum of carriers stands out, forming the foundation of the field of spintronics~\cite{vzutic2004spintronics,pulizzi2012spintronics,han2016perspectives,liu2020spintronics,yang2021chiral}. A key feature of spintronics is the ability to control spin angular momentum through the manipulation of spin-orbit interactions in materials with strong spin-orbit coupling. However, spintronics is inherently limited to materials with significant spin-orbit coupling. In recent years, growing interest has emerged in orbitronics, which leverages the orbital angular momentum (OAM) of electrons as a versatile degree of freedom. Unlike spintronics, orbitronics can operate even in systems without spin-orbit coupling~\cite{phong2019optically,go2021orbitronics,rappoport2023first,jo2024spintronics}, offering a broader range of applicability for advanced information processing and device functionality.

Although the concept of OAM has long been discussed in the literature to describe the orbital properties of materials, the field has experienced a revival over the past decade with the development of the modern theory of magnetization~\cite{thonhauser2005orbital,shi2007quantum}. This theory incorporates both inter-atomic and intra-atomic contributions to OAM. Describing electron motion in solids using wave packets, the self-rotation of these wave packets around their centre of mass gives rise to the concept of OAM.
OAM plays a crucial role in influencing various transport properties, including magnetoresistance in multivalley systems~\cite{zhou2019valley,das2021intrinsic,faridi2023magnetoresistance}, magnetoconductance in Weyl semimetals~\cite{knoll2020negative,sharma2020sign}, and chiral magnetic effects~\cite{ma2015chiral}. Similarly, analogous to the spin Hall effect~\cite{hirsch1999spin,schliemann2006spin,niimi2015reciprocal,sinova2015spin}, a longitudinal electric field can induce a transverse flow of OAM, resulting in the orbital Hall effect (OHE)~\cite{bernevig2005orbitronics,kontani2007study,kontani2008giant,tanaka2008intrinsic,kontani2009giant,jo2018gigantic,go2018intrinsic}.
Contrary to earlier beliefs that quenched OAM does not significantly affect transport phenomena, recent studies have shown that even in centrosymmetric crystals, with preserved inversion symmetry, OHE can arise via the transverse flow of nonequilibrium OAM induced by an electric field~\cite{go2018intrinsic}.

The first theoretical studies on the OHE focused on three-dimensional systems, such as $p$-doped silicon and transition metals~\cite{kontani2007study,kontani2008giant,tanaka2008intrinsic}. Recently, the promising potential of two-dimensional (2D) systems for applications in orbitronic-based devices has garnered significant attention~\cite{bhowal2021orbital,pezo2023orbital,cysne2024controlling,cysne2024transport,ghosh2023orbital,cysne2021disentangling,cysne2022orbital,pezo2022orbital,barbosa2024orbital,cysne2023ultrathin,chen2024topology,wang2024topological,gobel2024orbital}. Monolayers and bilayers of gapped graphene~\cite{bhowal2021orbital,pezo2023orbital,cysne2024controlling,cysne2024transport,ghosh2023orbital}, transition metal dichalcogenides (TMDs) \cite{cysne2021disentangling,cysne2022orbital,pezo2022orbital,barbosa2024orbital} and phosphorene~\cite{cysne2023ultrathin} are among the most notable candidates for exhibiting OHE in 2D materials.
In addition, several indirect confirmations of OHE have been achieved through experiments on orbital pumping~\cite{santos2023inverse,go2023orbital,hayashi2024observation} and orbital torque measurements~\cite{zheng2020magnetization,ding2020harnessing,go2020orbital,lee2021efficient,lee2021orbital}.  One breakthrough in the experimental measurement of OHE was reported in the magneto-optical observation of orbital accumulation at the surface of Ti and Cr~\cite{choi2023observation, lyalin2023magneto}. This orbital accumulation is directly related to the orbital Hall conductivity in the system which can be found theoretically.  The detection of inverse OHE in Ti and Mn~\cite{wang2023inverse} as well as the observation of ballistic orbital angular momentum currents in W~\cite{seifert2023time} are among other experimental realisations of the OHE. These findings have paved the way for substantial progress in the field of orbitronics.

Theoretical efforts to understand the physics behind the OHE have primarily focused on intrinsic mechanisms, where the geometrical features of the band structure play a central role. The Berry curvature acts as a momentum-space magnetic field, deflecting electrons carrying orbital angular momentum transversely. Intrinsic OHE is governed by the material's crystal structure and electronic properties, making it particularly significant in materials with broken inversion symmetry. In such systems, orbital angular momentum components align naturally due to band topology. Additionally, the intrinsic effect often dominates near the band edges (low Fermi energies), where the Berry curvature contribution is strongest. This mechanism is robust against external perturbations, such as impurities.
However, similar to the spin Halleffect, where extrinsic contributions are equally significant~\cite{inoue2004suppression,raimondi2005spin,khaetskii2006nonexistence,inoue2006vertex,milletari2017covariant,culcer2017interband}, extrinsic OHE plays a crucial role, particularly in real-world systems where disorder is unavoidable. Furthermore, discrepancies between theoretical predictions based on intrinsic mechanisms and experimental observations~\cite{choi2023observation} highlight the need for comprehensive studies on the role of disorder in OHE. 

To date, there have been limited investigations into the extrinsic contributions to OHE in 2D systems~\cite{liu2024dominance,canonico2024orbital,veneri2024extrinsic,tang2024role}. Most of these studies focus on inversion symmetry-broken gapped graphene models~\cite{liu2024dominance,canonico2024orbital,veneri2024extrinsic}, with one notable exception studying a centrosymmetric $p$-orbital triangular lattice model~\cite{tang2024role}.
In doped systems, it has been proposed  \cite{liu2024dominance} that extrinsic effects associated with the Fermi surface, such as skew scattering and side-jump, account for approximately 95\% of the OHE at experimentally relevant transport densities. This finding challenges the previous focus on intrinsic mechanisms and suggests that the OHE is primarily extrinsic in nature. While some studies emphasize on the dominance of extrinsic effects, others highlight the importance of intrinsic mechanisms in certain scenarios. For instance, in multiband ferromagnetic metals with dilute impurities, an extrinsic-to-intrinsic crossover occurs when the relaxation rate is comparable to the spin-orbit coupling \cite{PhysRevLett.97.126602}. Additionally, in 2D materials, both metallic and insulating phases can exhibit large orbital Hall currents, sometimes surpassing spin Hall currents \cite{PhysRevB.101.075429}.

The OHE in inversion-symmetric systems differs fundamentally from that in inversion symmetry-broken systems. In centrosymmetric materials, OAM cannot be assigned to individual bands due to the inversion symmetry. Consequently, the intrinsic OHE arises solely from interband transitions. However, when an electric field is applied, these interband transitions can generate orbital textures, leading to OHE.
In contrast, noncentrosymmetric systems exhibit intraband OAM, allowing both diagonal and off-diagonal density matrix contributions. These intraband components flow oppositely under the influence of an electric field, contributing to OHE. This fundamental difference raises an essential question: how does preserving or breaking the inversion symmetry influence the OHE?
While previous studies have investigated the impact of inversion symmetry breaking on the intrinsic OHE~\cite{PhysRevB.103.085113} in a simple cubic lattice with two atoms in the unit cell, the critical role of extrinsic contributions in this context remains largely unexplored. Moreover, the limited diversity of systems studied to date makes it challenging to draw definitive conclusions about the interplay between symmetry breaking and extrinsic mechanisms. In addition, the study of OHE in noncentrosymmetric materials also enriches our understanding of topological physics and expanding the range of candidate materials for topological applications.

In this paper, we aim to address this gap by studying a bilayer TMD system. The family of TMDs has been identified as a promising candidate for exhibiting strong OHE~\cite{bhowal2020intrinsic1,bhowal2020intrinsic,xue2020imaging}. While monolayers of TMDs are inherently noncentrosymmetric, bilayers consisting of two monolayers rotated by $\pi$ relative to each other restore inversion symmetry, making them centrosymmetric. This preserved inversion symmetry is advantageous for experimental observation of OHE, as it eliminates the possibility of confounding it with the valley Hall effect. It is well-known that the orbital and valley Hall effects are governed by fundamentally different symmetry principles. The valley Hall effect emerges only when inversion symmetry is broken, whereas the orbital Hall effect can occur regardless of such symmetry constraint. Therefore, in experiments, determining the source of edge magnetic moment accumulation under a longitudinal electric field is complex in inversion symmetry broken systems due to overlapping contributions from the VHE and OHE~\cite{cysne2021disentangling}. In fact, in real noncentrosymmetric materials with multiple bands involved, the intraband OAMs contribute in both VHE and OHE, but the interband terms only give a contribution in OHE. In contrast, centrosymmetric systems exhibit only the OHE, allowing for a more straightforward interpretation.
In this system, inversion symmetry can be easily broken by applying a gate voltage that induces a bias between the two layers. The strength of the gate voltage can be used to tune the degree of symmetry breaking. This feature makes bilayer TMDs a versatile platform for exploring OHE in both inversion-symmetric and inversion-asymmetric scenarios. In theoretical calculations, this symmetry breaking can be introduced by adding a term to the Hamiltonian that mimics the gate voltage, facilitating a controlled study of the OHE.

Using the effective Hamiltonian described in Refs.~\cite{cysne2022orbital,gong2013magnetoelectric,kormanyos2018tunable} and following a quantum kinetic theory approach~\cite{Vasko2005} we analyze both intrinsic and extrinsic contributions to the OHE in bilayer TMDs under unbiased (centrosymmetric) and biased (noncentrosymmetric) conditions. While the intrinsic OHE results in our work align precisely with previous findings using the Kubo formalism~\cite{cysne2022orbital}, our calculations highlight the significant role of extrinsic contributions, which cannot be neglected under any circumstances. Our key finding is that inversion symmetry breaking leads to a substantial enhancement of the extrinsic OHE. In centrosymmetric systems, the absence of intraband OAM restricts extrinsic contributions, as the diagonal part of the density matrix does not contribute to the extrinsic OHE. Nevertheless, even in these systems, extrinsic contributions become dominant gor larger Fermi energies. Additionally, we observe that in centrosymmetric systems, incorporating extrinsic terms can lead to a pronounced reduction and even a potential sign change in the total orbital Hall conductivity in hole-doped cases as the Fermi energy increases.

The rest of the paper is organized as follows. In Section~\ref{two}, the effective Dirac model Hamiltonian used to find the OHE is described and we have found the eigenvalues and eigenvectors for both biased and unbiased TMD bilayer. In Section \ref{three}, we describe the quantum kinetic formulation employed for the density matrix derivation in this work . Section \ref{four} addresses the OHE following the scheme where both intransite and intersite contributions to the OHE are taking into account. Our results for the centrosymmetric bilayer TMD (unbiased) and also its noncentrosymmetric counterpart (biased) are presented in this section. Finally, our results are summarized in section \ref{five} and some final remarks are presented. 

\section{Model hamiltonian for bilayer TMD}\label{two}


In spite of the fact that entire Brillouin zone has a contribution in orbital Hall conductivity, since the predominant contribution to the orbital moment comes from the $K$ and $K'$ valley points in TMDs ~\cite{bhowal2021orbital,cysne2021disentangling,pezo2022orbital,bhowal2020intrinsic1,bhowal2020intrinsic}, we have concentrated on the contributions around these points. This approach is consistent with previous calculations on both intrinsic~\cite{cysne2021disentangling,cysne2022orbital} and extrinsic~\cite{liu2024dominance,veneri2024extrinsic} orbital Hall effects, which have also employed a low-energy Hamiltonian around the valleys which not only makes these systems computationally tractable, bu also enables a more efficient analysis without sacrificing essential physical insights.
Following Ref.~\cite{cysne2022orbital}, we use an effective Hamiltonian to describe the conduction and valence bands near $ K$ and $ K'$ points of an unbiased bilayer TMD. Here we consider the case of 2H-MoS$_2$ as the prototypical member of the vast class of TMDs. In the tight binding basis $\beta_{tb}=\lbrace \ket{d^1_{z^2}},\frac{1}{\sqrt{2}}(\ket{d^1_{x^2-y^2}}-i\tau\ket{d^1_{xy}}),\ket{d^2_{z^2}},\frac{1}{\sqrt{2}}(\ket{d^2_{x^2-y^2}}+i\tau\ket{d^2_{xy}})  \rbrace$ with the superscript 1 and 2 illustrative of two layers, the effective low-energy Hamiltonian at zero temperature for unbiased case is given by
\begin{equation}\label{Hm}
{\cal H}_0(\bm k)=\begin{pmatrix}
2m&\gamma_+&0&0 \\ \gamma_- &-\tau s_z\lambda&0&t_{\bot}\\0&0&2m&\gamma_-\\0&t_{\bot}&\gamma_+&\tau s_z\lambda
\end{pmatrix},
\end{equation}
where $\tau=\pm$ is the valley index, $s_z$ is the Pauli matrix in the spin space, $\gamma_{\pm}=\hbar v_{\rm F}(\tau k_x\pm ik_y)$ and $\hbar v_{\rm F}=at$ with $a=3.160\,\rm{\AA}$ the lattice constant, $t=1.137\,\rm{eV}$ the intralayer hopping, $2m=1.766\,\rm{eV}$ the band gap, $t_{\bot}=0.043\,\rm{eV}$ the interlayer hopping and $\lambda=0.073\,\rm{eV}$ the spin-orbit coupling. Since the OHE in doped case is weakly affected by spin-orbit coupling in TMDs, we consider spin-degenerate bands and set $\lambda=0$ in our calculations\cite{cysne2021disentangling,cysne2022orbital}. In this Hamiltonian, each layer is described by an effective Dirac model and an interlayer hopping term connects the two layers. The eigenvalues of the system are given by
\begin{eqnarray}
&&{\ket{\phi_{1,v,\tau}}=\frac{1}{\sqrt{2}}\Bigl(\tau\eta_{k,1}^{+}e^{-i\tau\varphi}\!,-\zeta_{k,1}^{+}\!,-\tau\eta_{k,1}^{+}e^{i\tau\varphi}\!,\zeta_{k,1}^{+} \Bigr)^T\!\!\!\!,}\qquad\label{evv1}\\
&&\ket{\phi_{2,v,\tau}}=\frac{1}{\sqrt{2}}\Bigl(-\tau\eta_{k,2}^{+}e^{-i\tau\varphi}\!,\zeta_{k,2}^{+},-\tau\eta_{k,2}^{+}e^{i\tau\varphi}\!,\zeta_{k,2}^{+} \Bigr)^T\!\!\!\!,\label{evv2}
\end{eqnarray}
for the valence bands ($v$) and for the conduction bands ($c$) we have
\begin{eqnarray}
&&{\ket{\phi_{1,c,\tau}}=\frac{1}{\sqrt{2}}\Bigl(\tau\eta_{k,1}^{-}e^{-i\tau\varphi}\!,-\zeta_{k,1}^{-}\!,-\tau\eta_{k,1}^{-}e^{i\tau\varphi}\!,\zeta_{k,1}^{-} \Bigr)^T\!\!\!\!,}\qquad\\\label{evv3}
&&{\ket{\phi_{2,c,\tau}}=\frac{1}{\sqrt{2}}\Bigl(-\tau\eta_{k,2}^{-}e^{-i\tau\varphi}\!,\zeta_{k,2}^{-},-\tau\eta_{k,2}^{-}e^{i\tau\varphi}\!,\zeta_{k,2}^{-} \Bigr)^T\!\!\!\!,}\label{evv4}
\end{eqnarray}

where $\eta_{k,i}^{\pm}=\sqrt{1-(\zeta_{k,i}^{\pm})^2}$, $\zeta_{k,i}^{\pm}=\frac{m_i+\lambda_{k,i}^{\pm}}{\sqrt{(m_i+\lambda_{k,i}^{\pm})^2+(\hbar v_{\rm F}k)^2}}$, $\lambda_{k,i}^{\pm}=\pm\sqrt{{m_i^2+(\hbar v_{\rm F}k)^2}}$, $m_{1(2)}=m\pm\allowbreak t_{\bot}/2$ and $\varphi$ being the angle between $\bm k$ and $\hat{x}$. Then the eigenvalues take the form
\begin{eqnarray}
&&\varepsilon_{v,1(2)}(k)=m_{2(1)}-\lambda_{k,1(2)}^{+},\\
&&\varepsilon_{c,1(2)}(k)=m_{2(1)}-\lambda_{k,1(2)}^{-}.
\end{eqnarray} 
The eigenvalues and eigenvectors of the system found from a perturbative scheme~\cite{cysne2022orbital} are also presented in Appendix~\ref{ap-one}.

\begin{figure}[ht]
 \captionsetup[subfigure]{labelformat=empty}
\centering
  \subfloat[]{\includegraphics[width=0.45\linewidth]{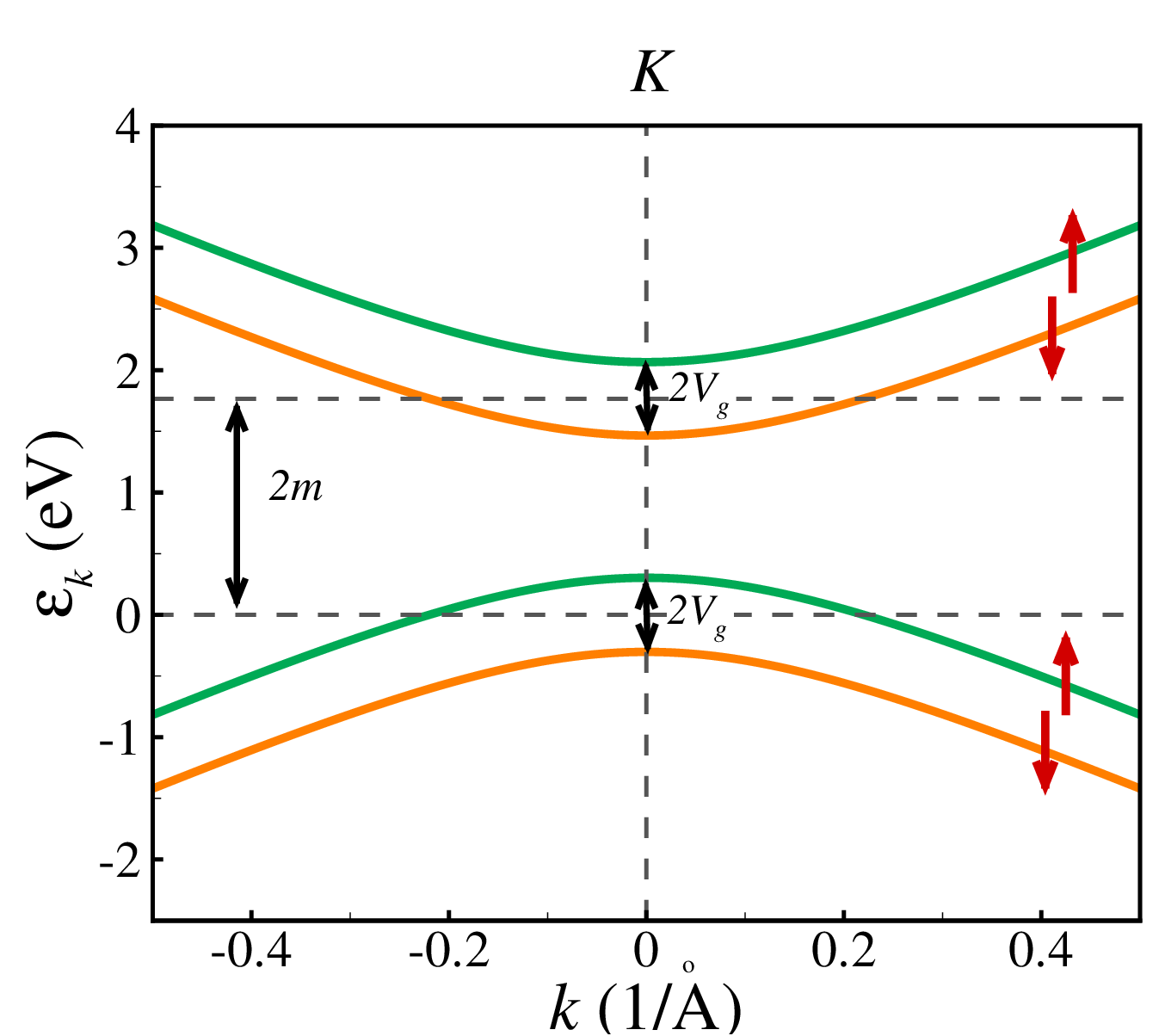}}\negthickspace
\hfil
    \subfloat[]{\includegraphics[width=0.45\linewidth]{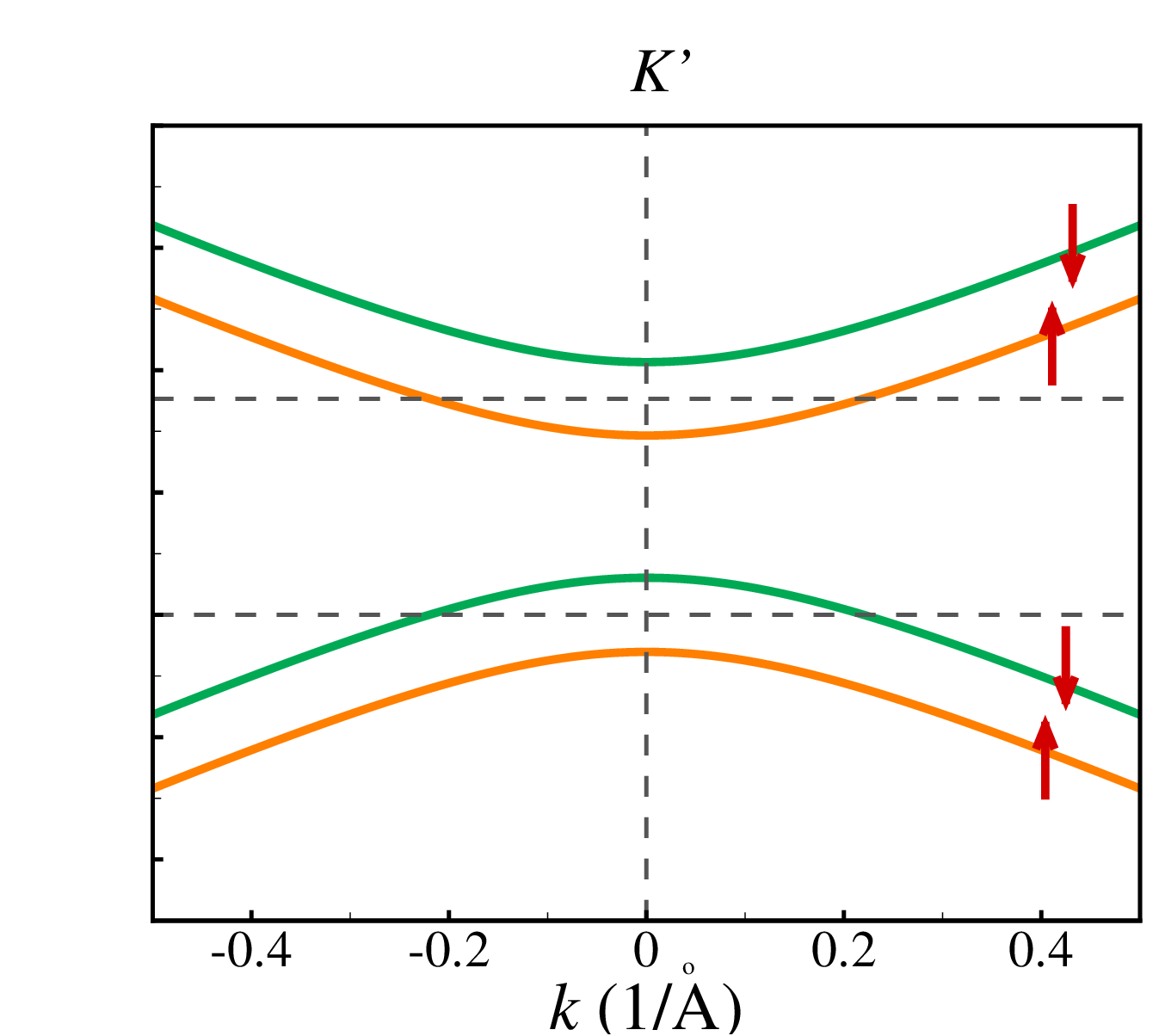}}

\caption{(Color online) The energy spectrum of the biased bilayer of 2H-MoS$_2$ with energy gap $2m$ near $K$ and $K'$ valley for $V_g=0.3$ eV. The green and the orange lines belong to layer 1 and 2 respectively (Eqs.~\eqref{eev} and \eqref{eec}) and the red arrows show the orbital magnetic moment in each band. Note that zero energy is defined at the valence band maximum when the gate voltage is absent.  \label{fig1}    }
 \end{figure}

We are also interested to go further and break the inversion symmetry in the system and make predictions about the OHE in this case. In noncentrosymmetric materials, the intrinsic orbital moment is already present, leading to a more robust OHE. Therefore, the mechanism is fundamentally different and may make noncentrosymmetric solids more suitable for direct observation of the OHE. Previously, by making use of a tight-binding model Hamiltonian of a simple cubic lattice with two atoms in the unit cell, the effect of the inversion symmetry breaking on the OHE was studied \cite{PhysRevB.103.085113}. They showed that with the increase in the strength of the broken symmetry, the magnitude of the intrinsic orbital moment in the Brillouin zone increases. Here we focus on a real 2D system. This can be achieved by including a gate voltage $V_g$ in the Hamiltonian of bilayer TMD such that
\begin{equation}\label{HU}
{\cal H}_0^{V_g}(\bm k)=\begin{pmatrix}
2m+V_g&\gamma_+&0&0 \\ 
\gamma_- &+V_g&0&t_{\bot}\\0&0&2m-V_g&\gamma_-\\0&t_{\bot}&\gamma_+&-V_g
\end{pmatrix}.
\end{equation}
Following Ref.~\cite{cysne2022orbital}, the interlayer hopping is considered as a perturbation, but  in the first order in $t_{\bot}$, no correction is made due to the hopping. Therefore the eigenvectors of the unperturbed Hamiltonian ($t_\bot=0$ in Eq.~\eqref{HU}) are the eigenvectors of the system in this case and since the eigenvectors are not affected by the gate voltage, they are the same as the eigenstates of the unbiased unperturbed system which are given by
\begin{eqnarray}
&&\ket{\psi_{1,v,\tau}}=\frac{1}{\sqrt{2}}\Bigl(-\tau\xi_k^{-}e^{-i\tau\varphi},\xi_k^{+},0,0 \Bigr)^T,\label{uev1}\\
&&\ket{\psi_{2,v,\tau}}=\frac{1}{\sqrt{2}}\Bigl(0,0,-\tau\xi_k^{-}e^{i\tau\varphi},\xi_k^{+}\Bigr)^T,\label{uev2}
\end{eqnarray}
for the valence bands and
\begin{eqnarray}
&&\ket{\psi_{1,c,\tau}}=\frac{1}{\sqrt{2}}\Bigl(\tau\xi_k^{+}e^{-i\tau\varphi},\xi_k^{-},0,0 \Bigr)^T,\label{uev3}\\
&&\ket{\psi_{2,c,\tau}}=\frac{1}{\sqrt{2}}\Bigl(0,0,\tau\xi_k^{+}e^{i\tau\varphi},\xi_k^{-}\Bigr)^T,\label{uev4}
\end{eqnarray}
for the conduction bands in each layer. Here $\xi_k^{\pm}=\sqrt{1\pm\xi_k}$ and $\xi_k=m/\lambda_k$ with $\lambda_k=\allowbreak\sqrt{(\hbar v_{\rm F}k)^2+m^2}$. On the other hand, while the gate voltage makes an upward shift in the energy dispersion of conduction and valence bands in layer 1, it causes  a downward shift in that of layer 2 such that the degeneracy of the bands is lifted. In this case the eigenvalues take the form 
\begin{eqnarray}
&&\varepsilon_{v,1(2)}(k)=m\pm V_g-\lambda_k,\label{eev}\\
&&\varepsilon_{c,1(2)}(k)=m\pm V_g+\lambda_k.\label{eec}
\end{eqnarray} 
It is important to note that in presence of a gate voltage $V_g$, this perturbative scheme is valid for small values of $t_{\bot}/V_g$. A comparison between the exact and perturbative eigenvectors (see Appendix~\ref{ap-two}) show that for the certain value of $t_{\bot}=0.043$ eV, the perturbative scheme works well for $V_g\geqslant 0. 2$ eV. In order to study the biased bilayer for smaller values of the gate voltage and the values comparable to the interlayer coupling, a fully numerical approach is required. The energy dispersion of the bare Hamiltonian, ${\cal H}_0^{V_g}$ is shown in Fig.~\ref{fig1}  for both electron and hole-doped bilayer 2H-MoS$_2$ and the contribution of the orbital magnetic moment is determined by arrows.

\section{Quantum Kinetic equation and linear response}\label{three}
The transport theory we employ here is based on a density matrix approach which accounts for disorder and electric field-corrections in the collision integral~\cite{culcer2017interband,atencia2022semiclassical} which has been widely used in literature~\cite{Vasko2005,farokhnezhad2022spin,mehraeen2024quantum}. We consider the total Hamiltonian of the system of the form ${\cal H}={\cal H}_0+V(\bm r)+U(\bm r)$ where ${\cal H}_0$ is the effective bare band Hamiltonian referring to whether biased or unbiased case, $V(\bm r)=e\bm E\cdot\bm r$ represents the electric field-induced perturbation and $U(\bm r)$ is the disorder potential. Starting from Liouville's equation for the single-particle density operator $\frac{\partial\rho}{\partial t}+\frac{i}{\hbar}[{\cal H},\rho]=0$ and writing the density matrix as $\rho=\langle\rho\rangle+g_0$ with $\langle\rho\rangle$ and $g_0$ being the averaged over disorder configurations and the fluctuating parts respectively, we obtain the following quantum kinetic equation for $\langle\rho\rangle$ 
\begin{equation}\label{kinetic}
\frac{\partial\langle\rho\rangle}{\partial t}+\frac{i}{\hbar}[{\cal H}_0,\langle\rho\rangle]+J_0(\langle\rho\rangle)=-\frac{i}{\hbar}[V,\langle\rho\rangle]-J_E(\langle\rho\rangle),
\end{equation}
where $J_0(\langle\rho\rangle)$ is the collision integral and is given by $J_0(\langle\rho\rangle)=\frac{i}{\hbar}\langle[U,g_0]\rangle$ with $g_0=-\frac{i}{\hbar}\times\allowbreak\int_0^\infty dt'[e^{-i{\cal H}_0t'/\hbar}Ue^{i{\cal H}_0t'/\hbar},\langle\rho(t)\rangle]$. $J_E(\langle\rho\rangle)=\frac{i}{\hbar}\langle[U,g_E]\rangle$ is also an electric field-mediated collision integral with $g_E=-\frac{i}{\hbar}\int_0^\infty dt''e^{-i{\cal H}_0t''/\hbar}[V,g_0(t-t'')]e^{i{\cal H}_0t''/\hbar}$ being a correction to the fluctuating part of the density matrix, $g_0$, due to the electric field.

Expressing Eq.~\eqref{kinetic} in crystal momentum representation, we write $f_k$ instead of $\langle\rho\rangle$ as the density matrix of the system. The first term on the RHS of the Eq.~\eqref{kinetic} can also be written as $\frac{e\bm E}{\hbar}\cdot\frac{Df_k}{D\bm k}$ in this representation where $\frac{Df_k}{D\bm k}=\frac{\partial f_k}{\partial\bm k}-i[\bm R_k,f_k]$ is the covariant derivative and $\bm R_k$ is the momentum space Berry connection and its elements between the band eigenstates are given by $\bm{R}_k^{mm'}=i\bra{u_k^m}\nabla_{\bm{k}} u_k^{m'}\rangle$.

In order to solve this equation, the density matrix is decomposed as $f_k=n_k+S_k$ where $n_k$ is diagonal in the band index and expresses the distribution of carriers in each band while $S_k$ is off-diagonal and represents the band-mixing. On the other hand, the equilibrium density matrix is diagonal and its elements are the Fermi-Dirac distribution function for each band. In presence of an external electric field, the density matrix can be written as $f_k^{mm^\prime}=f_0\left(\varepsilon_k^m\right)\delta_{mm^\prime}+f_{Ek}^{mm^\prime}$ with $f_{Ek}^{mm^\prime}=n_{Ek}^m\delta_{mm^\prime}+S_{Ek}^{mm^\prime}$ being the correction to the equilibrium density matrix $f_0\left(\varepsilon_k^m\right)$ to the first order in the electric field. Then we have
\begin{equation}\label{kineticf}
\frac{\partial f_k}{\partial t}+\frac{i}{\hbar}[{\cal H}_0,f_k]+J_0(f_k)=\frac{e\bm E}{\hbar}\cdot\frac{Df_k}{D\bm k}-J_E(f_k).
\end{equation} 
Using the Born approximation for the collision integrals, the density matrix can be written as an expansion in the impurity density $n_i$. This is because for a system with randomly distributed impurities, the disorder potential takes the form $U(\bm r)=\sum_{i}u(\bm r-\bm{r}_i)$ with $u$ the potential of a single impurity and $\bm{r}_i$ the impurity locations. Using short-range disorder rather than long-range disorder has several advantages, especially when analyzing the extrinsic contributions to OHE. To create more manageable computer models and simpler mathematical expressions, short-range disorder might be represented as point-like contaminants with a localized interaction range. Localized scattering events are predominant in extrinsic OHE mechanisms. These processes including side-jump, a transverse displacement during scattering, and skew scattering, an asymmetric scattering process that produces a net transverse orbital current, are directly influenced by short-range disorder. Furthermore, screening effects have less of an impact on short-range disorder, making its influence on OHE more apparent. On the other hand, short-range impurities are the most common kind of disorder in many materials found in the actual world. Therefore, research on short-range disorder is more in line with device applications and experimental findings.  However, long-range disorder, leads to smoother variations in the potential which primarily affects the band structure broadening and less the localized scattering mechanisms, and it is more relevant for phenomena like mobility degradation but is less directly tied to the extrinsic mechanisms that drive OHE.

While the first order term in the potential vanishes due to the randomness of the impurities, the second order term in the Bloch representation takes the form $\langle U_{kk'}^{mm'}U_{k'k}^{m''m'''}\rangle=\allowbreak n_iu_{kk'}^{mm'}u_{k'k}^{m''m'''}$. Back to the collision integrals, it can be shown that the leading order for the band-diagonal density matrix $n_{Ek}$ is -1 while the off-diagonal $S_{Ek}$ does not depend on the impurity concentration and starts at order 0. Therefore for the diagonal part to the leading order ${-1}$, Eq.~\eqref{kineticf} reduces to $[J_0(n_E^{(-1)})]_k^m=\frac{e\bm E}{\hbar}\cdot\frac{\partial f_0(\varepsilon_k^m)}{\partial \bm k}$  where the Born approximation collision integral is $[J_0(n_E)]_k^m=\frac{2\pi}{\hbar}\sum_{m',k'}\langle U_{kk'}^{mm'}U_{k'k}^{m'm}\rangle(n_{Ek}^m-n_{Ek'}^{m'})\delta(\varepsilon_k^m-\varepsilon_{k'}^{m'})$. For a system with isotropic band dispersions, the solution is
\begin{equation}
n_{Ek}^{m(-1)}=\tau^m_p\frac{e\bm E}{\hbar}\cdot\frac{\partial\varepsilon_k^m}{\partial \bm k}\frac{\partial f_0(\varepsilon_k^m)}{\partial \varepsilon_k^m},
\end{equation}
where ${\frac{1}{\tau^m_p}=\frac{2\pi}{\hbar}\sum_{m',k'}\langle U_{kk'}^{mm'}U_{k'k}^{m'm}\rangle[1-\cos(\varphi'-\varphi)]}\delta(\varepsilon_k^m-\varepsilon_{k'}^{m'})$ is the transport time. The solution of Eq.~\eqref{kineticf} for $S_{Ek}$ in zeroth order to $n_i$ takes the form
\begin{equation}\label{S}
S_{Ek}^{(0)mm'}=\frac{\hbar(D+D')_{Ek}^{mm'}}{i(\varepsilon_k^m-\varepsilon_k^{m'}-i\eta)},
\end{equation}
which is composed of two off-diagonal contributions. The first one which is the intrinsic term, expresses the response of the Fermi sea and is given by
\begin{equation}\label{D}
D_{Ek}^{mm'}=\frac{ie}{\hbar}\bm E\cdot\bm R_k^{mm'}[f_0(\varepsilon_k^m)-f_0(\varepsilon_k^{m'})].
\end{equation}
The second term is an anomalous extrinsic driving term found from the off-diagonal elements of the collision integral ${D'}_{Ek}^{mm'}=-[J_0(n_E^{(-1)})]_k^{mm'}$ and can be considered as a Fermi surface contribution to the side jump given by
\begin{equation}\label{D'}
{D'}_{Ek}^{mm'}=-\frac{\pi}{\hbar}\sum_{m'',k'}\langle U_{kk'}^{mm''}U_{k'k}^{m''m'}\rangle
\bigl\lbrace(n_{Ek}^{m'}-n_{Ek'}^{m''})\delta(\varepsilon_k^{m'}-\varepsilon_{k'}^{m''})+(n_{Ek}^{m}-n_{Ek'}^{m''})\delta(\varepsilon_{k'}^{m''}-\varepsilon_{k}^{m})\bigr\rbrace.
\end{equation}
Since $S_{Ek}$ is found up to the zeroth order, $n_{Ek}$ should also be written up to subleading order. Using the equation $[J_0(n_E^{(0)})]_k^{m}=-[J_E(f_0)]_k^{m}-[J_0(S_E^{(0)})]_k^{m}$, we will have $n_E^{(0)}=n_{E,sj}^{(0)}+n_{E,sk}^{(0)}$ where $n_{E,\rm{sj}}^{(0)}=-\tau_p^m [J_E(f_0)]_k^{m}$ is another contribution of the side jump mediated by the electric field correction to the collision integral and $n_{E,sk}^{(0)}=-\tau_p^m [J_0(S_E^{(0)})]_k^{m}$ is the skew scattering contribution with

\begin{equation}\label{sj}
\begin{split}
[J_E(f_0)]_k^{m}&=\frac{2\pi}{\hbar}\frac{\partial f_0(\varepsilon_k^m)}{\partial\varepsilon_k^m}e\bm E\cdot\sum_{m'\bm k'}\langle U_{kk'}^{mm'}U_{k'k}^{m'm}\rangle[\bm R_{k'}^{m'm'}-\bm R_{k}^{mm}]\delta(\varepsilon_{k'}^{m'}-\varepsilon_{k}^{m})\\
&+\frac{2\pi}{\hbar}\frac{\partial f_0(\varepsilon_k^m)}{\partial\varepsilon_k^m}e\bm E\cdot\sum_{m'\bm k'}\mathrm{Im}\bigl\lbrace\langle[(\nabla_{\bm k}+\nabla_{\bm {k'}}) U_{kk'}^{mm'}]U_{k'k}^{m'm}\rangle\bigr\rbrace\delta(\varepsilon_{k'}^{m'}-\varepsilon_{k}^{m}),
\end{split}
\end{equation}
and
\begin{equation}\label{sk}
\begin{split}
[J_0(S_E^{(0)})]_k^{m}=-\frac{2\pi^2}{\hbar}\sum_{m'm''n\bm k'\bm k''} &
\biggl\lbrace  \mathrm{Im}\biggl[\frac{\langle U_{kk'}^{mm''}U_{k'k}^{m'm}\rangle\langle U_{k'k''}^{m''n}U_{k''k'}^{nm'}\rangle}{(\varepsilon_{k'}^{m''}-\varepsilon_{k'}^{m'})}\biggr]{ K}^{n m  m{'}m{''}}_{k k{'} k{''}}\delta(\varepsilon_{k'}^{m''}-\varepsilon_{k}^{m})\\
&-\mathrm{Im}\biggl[\frac{\langle U_{kk'}^{m''m'}U_{k'k}^{m'm}\rangle\langle U_{kk''}^{mn}U_{k''k}^{nm''}\rangle}{(\varepsilon_{k}^{m}-\varepsilon_{k}^{m''})}\biggr]  { K}^{n m  m{''}m{}}_{k k k{''}}  \delta(\varepsilon_{k'}^{m'}-\varepsilon_{k}^{m''})\biggr\rbrace,
\end{split}
\end{equation}
where
\begin{equation}\label{kk}
{K}^{n m  ij}_{k l p}=(n_{El}^{i(-1)}-n_{Ep}^{n(-1)})\delta(\varepsilon_{l}^{i}-\varepsilon_{p}^{n})+(n_{El}^{j(-1)}-n_{Ep}^{n(-1)})\delta(\varepsilon_{p}^{n}-\varepsilon_{l}^{j}).
\end{equation}

As can be seen From Eqs.~\eqref{sk} and ~\eqref{kk}, the skew scattering term is third order in the electron-impurity potential. It is worth noting that the parameter $n_i$ cancels out in the skew scattering and side jump equations, making the OHE independent of the specifics of short-range impurities.  
\section {Orbital Hall effect}\label{four}
In linear response theory, the orbital Hall current is a transverse response in the $\hat y$ direction mediated by the $\hat{z}$ component of the OAM operator in response to a longitudinal electric field pointed to $\hat{x}$ and in the language of density matrix theory it is given by $j_y=\mathrm{Tr}[\hat{\rho}\,\hat{\bm{J}}_y]$ where $\hat{\rho}$ is the density matrix and $\hat{\bm{J}}_y=\frac{1}{2}\lbrace\hat{L}_z,\hat{v}_y\rbrace$ is the orbital Hall current operator. Here $\hat{v}_y$ is the transverse component of the velocity operator whose matrix elements in band representation are given by
\begin{equation}
\hat{\bm v}_{\bm{k}}^{mm'}=\bigl[\frac{i}{\hbar}[{\cal H}_0,\hat{\bm r}]\bigr]_k^{mm'}=\frac{1}{\hbar}\frac{\partial\varepsilon_k^m}{\partial\bm{k}}\delta_{m,m'}+\frac{i}{\hbar}(\varepsilon_k^m-\varepsilon_k^{m'})\bm{R}_k^{mm'}
\end{equation}
with $\hat{\bm r}_{\bm{k}}^{mm'}=i\partial_{\bm k}\delta_{m,m'}+\bm{R}_k^{mm'}$ being the position operator. Based on the modern theory of OAM, we use an expression for the OAM operator $\hat{L}_z$ which contains both intersite and intrasite contributions to the OAM and applies the Berry phase of Bloch bands to define the orbital magnetic moment (OMM). In this description, we have $\hat{L}_z=\frac{-\hbar}{g_{\rm L}\mu_{\rm B}}\hat{\bm M}_z$ where $g_{\rm L}=1$ is the Land\'e g-factor, $\mu_{\rm B}=\frac{e\hbar}{2m_e}$ is the Bohr magneton and $m_e$ is the electron mass. $\hat{\bm M}_z$ is also the operator of the OMM which is connected to the Berry phase and is given by~\cite{bhowal2021orbital} 
\begin{equation}\label{m}
\hat{{\bm M}}=-\frac{e}{4}(\hat{\bm r}\times \hat{\bm v}-\hat{\bm v}\times \hat{\bm r})
\end{equation}
 and has only a component parallel to $\hat z$ in two-dimensional systems.
 
 \subsection {Centrosymmetric bilayer TMD}
From the definition of OMM, it follows that in the presence of both time-reversal and inversion symmetries, the OMM vanishes. Therefore, for the nearly degenerate bands of the unbiased bilayer transition metal dichalcogenide (TMD), and following Ref.~\cite{cysne2022orbital}, we adopt an expression for $\hat {\bm M}_z$ with a non-Abelian structure. In this formulation, the diagonal elements (representing the OMM in each band) are zero, while the off-diagonal elements remain nonzero. On the basis of Bloch bands or the eigenvectors of the unbiased system~\eqref{evv1}-\eqref{evv4}, the OMM operator is given by
\begin{equation}
\hat{{\bm M}}_z(k)=\begin{pmatrix}
\hat{M}^v_z(k)&\hat{0}_{2\times2}\\ \hat{0}_{2\times2}&\hat{M}^c_z(k)
\end{pmatrix},
\end{equation}
where $\hat{0}_{2\times2}$ is the $2\times2$ zero matrix and $\hat{M}^v_z(k)$ and $\hat{M}^c_z(k)$ are calculated in the subspace of valence and conduction respectively and are given by ${\hat{M}^{v(c)}_z(k)=\tau M^{v(c)}(k)\begin{pmatrix}
0&1\\ 1&0\end{pmatrix}}$ with $M^{v(c)}(k)$ calculated numerically. 
\begin{figure}[h]
\centering
  \includegraphics[width=4.in]{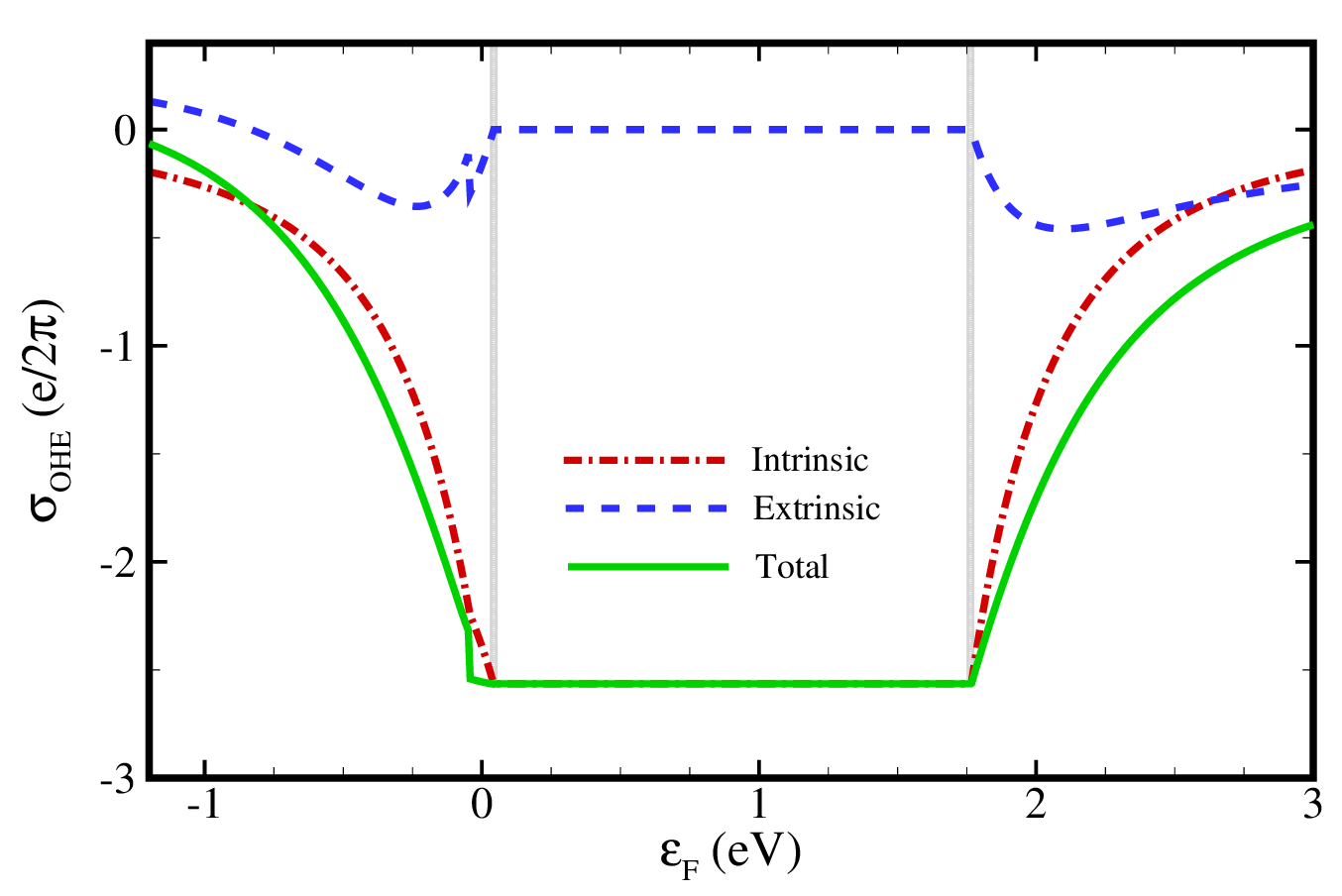}
\caption{\label{fig2} (Color online) Intrinsic, extrinsic and the total orbital Hall conductivity $\sigma_{\rm{OHE}}$ versus the Fermi energy $\varepsilon_{\rm F}$ for unbiased centrosymmetric bilayer of 2H-MoS$_2$. The vertical grey lines delimit the band gap. The extrinsic conductivity predominates over the intrinsic one as we increase the Fermi energy in electron-dopes system and it experiences a sign change for the hole-doped case. 
    }
\end{figure}
Here, in contrast to the inversion asymmetric systems, the diagonal part of the OMM operator (and through that the OAM operator) which expresses the intraband elements of the OMM vanishes as expected from a centrosymmetric system (Note that the diagonal components of $\hat{M}^{v(c)}_z$ are zero , and the same holds for $\hat{{\bm M}}_z$). This has important consequences in the extrinsic orbital Hall conductivity in the system. It leads to zero diagonal elements in both the OAM operator $\hat{L}_z$ and also the orbital Hall current operator $\hat{\bm J}_y$. As the orbital Hall conductivity is defined as $j_y=\mathrm{Tr}[\hat{\rho}\,\hat{\bm J}_y]$, the diagonal part of the density matrix, $\hat{n}_{Ek}$, has no contribution in the orbital Hall conductivity and we are left with the off-diagonal elements of the density matrix, $\hat{S}_{Ek}$. Therefore the extrinsic skew scattering (Eq.~\eqref{sk}) and the electric field induced contribution of the side-jump (Eq.~\eqref{sj}) are absent for the unbiased  TMD bilayer and the only contribution of the extrinsic mechanisms stems from the zeroth order off-diagonal density matrix which can be found from Eqs.~\eqref{S} and \eqref{D'}. On the other hand, the intrinsic orbital Hall conductivity is given by substituting Eq.~\eqref{D} back in Eq.~\eqref{S}. We have numerically found the intrinsic orbital Hall conductivity for both hole-doped and electron-doped cases as can be seen in Fig.~\ref{fig2} which is exactly the same as previous calculations~\cite{cysne2022orbital}.  We can also find an analytical expression for the intrinsic part using the perturbative scheme (Appendix \ref{ap-one}) which to the linear order in $t_{\bot}/m$ is given by 
\begin{equation}\label{int0}
\sigma_{\rm{OHE}}^{\rm{int}}=-\sum_i\frac{g_sg_v}{4\hbar^2}\frac{em_e}{2\pi}\frac{m^2(\hbar v_{\rm F})^2}{\lambda_{k_{\rm F}^i}^3}\bigl[\frac{2}{3}+\gamma_i\frac{t_{\bot}}{2\lambda_{k_{\rm F}^i}}(\frac{1}{4}-\frac{m}{5\lambda_{k_{\rm F}^i}})\bigr]
\end{equation}
where the sum is over the bands intersecting the Fermi energy, ${k_{\rm F}^i}$ representing Fermi wave-vector of band $i$ and $\gamma_i=1$ for the bands $(v,+)$ and $(c,-)$ while $\gamma_i=-1$ for the bands $(v,-)$ and $(c,+)$ given by Eqs.~\eqref{ev1}-\eqref{ev4}. $g_s$ and $g_v$ are the spin and valley degeneracies, respectively, and each one is 2 in the studied system. We recall that the energy gap is $2m$ and $\lambda_{k_{\rm F}^i}=\sqrt{(\hbar v_{\rm F}{k_{\rm F}^i})^2+m^2}$. Putting ${k_{\rm F}^i}=0$ in electron-doped or hole-doped case, the in-gap orbital Hall conductivity is found as
\begin{equation}\label{ingap}
\sigma_{\rm{OHE}}^{\rm{in-gap}}=-\frac{g_sg_v}{3\hbar^2}\frac{em_e}{2\pi}\frac{(\hbar v_{\rm F})^2}{m}
\end{equation}
in agreement with previous calculations~\cite{cysne2022orbital}. The extrinsic orbital Hall conductivity is also found numerically using Eqs.~\eqref{S} and~\eqref{D'}. The anomalous extrinsic contribution which is a Fermi surface contribution to the side-jump is also plotted in Fig.~\ref{fig2} as a function of Fermi energy for both electron and hole-doped bilayer 2H-MoS$_2$.  Although in contrast to the intrinsic orbital Hall conductivity, the extrinsic contribution is small clost to the band gap, but as we move away from the gap the extrinsic conductivityincreases and finally dominates the intrinsic one in both hole-doped and electron-doped systems. On the other hand, interestingly the extrinsic contribution for the Fermi energy in the valence band changes sign at a certain Fermi energy which leads to a substantial deacrease of the total orbital Hall conductivity.

\subsection {Noncentrosymmetric bilayer TMD}
In the case of biased bilayer, the inversion symmetry is broken and the OMM can be found from the conventional expression ~\eqref{m}. On the basis of non-degenerate eigenvectors $\beta_{\psi}=\allowbreak\lbrace\psi_{1,c};\psi_{1,v} ;\psi_{2,c} ;\psi_{2,v}  \rbrace$ given by Eqs~\eqref{uev1}-\eqref{uev4}, the OMM operator in this case is given by 
\begin{equation}\label{mm}
{\hat{{\bm M}}}_z^{V_g}\left(k\right)=\tau M_0\left(k\right)\begin{pmatrix}
\hat{I}&\hat{0}_{2\times2}\\ \hat{0}_{2\times2}&-\hat{I}
\end{pmatrix},
\end{equation}
with $\hat{I}$ being the $2\times2$ identity matrix and $M_0(k)=\frac{e}{2\hbar}\frac{(\hbar v_{\rm F})^2m}{((\hbar v_{\rm F}k)^2+m^2)}$ being the OMM in massive Dirac model. As expected, here we have finite intraband terms. Note that in systems with broken inversion symmetry, off-diagonal terms need not vanish in general, but in this particular case, they do~\cite{bhowal2021orbital}.The orbital Hall current operator $\hat{\bm J}_y^{V_g}$ in this case takes the following form
\begin{equation}
{\hat{{\bm J}}}_y^{V_g}\left(k\right)=\begin{pmatrix}
{\hat{J}}_y^-&\hat{0}_{2\times2}\\ \hat{0}_{2\times2}&{\hat{J}}_y^+
\end{pmatrix},
\end{equation}
where ${\hat{J}}_y^{\pm}=v_{\rm F}\sin{\varphi}(\sqrt{1-\xi_k^2}\hat{\sigma_z}+\xi_k\hat{\sigma_x})\pm \tau v_{\rm F}\cos{\varphi}\hat{\sigma}_y$. For the biased case, all the intrinsic and extrinsic terms are present and similar to the gapped graphene model, after a lengthy algebra they can be found analytically as follows. 
\begin{figure}[h]
\centering
\subfloat{%
  \includegraphics[width=4.in]{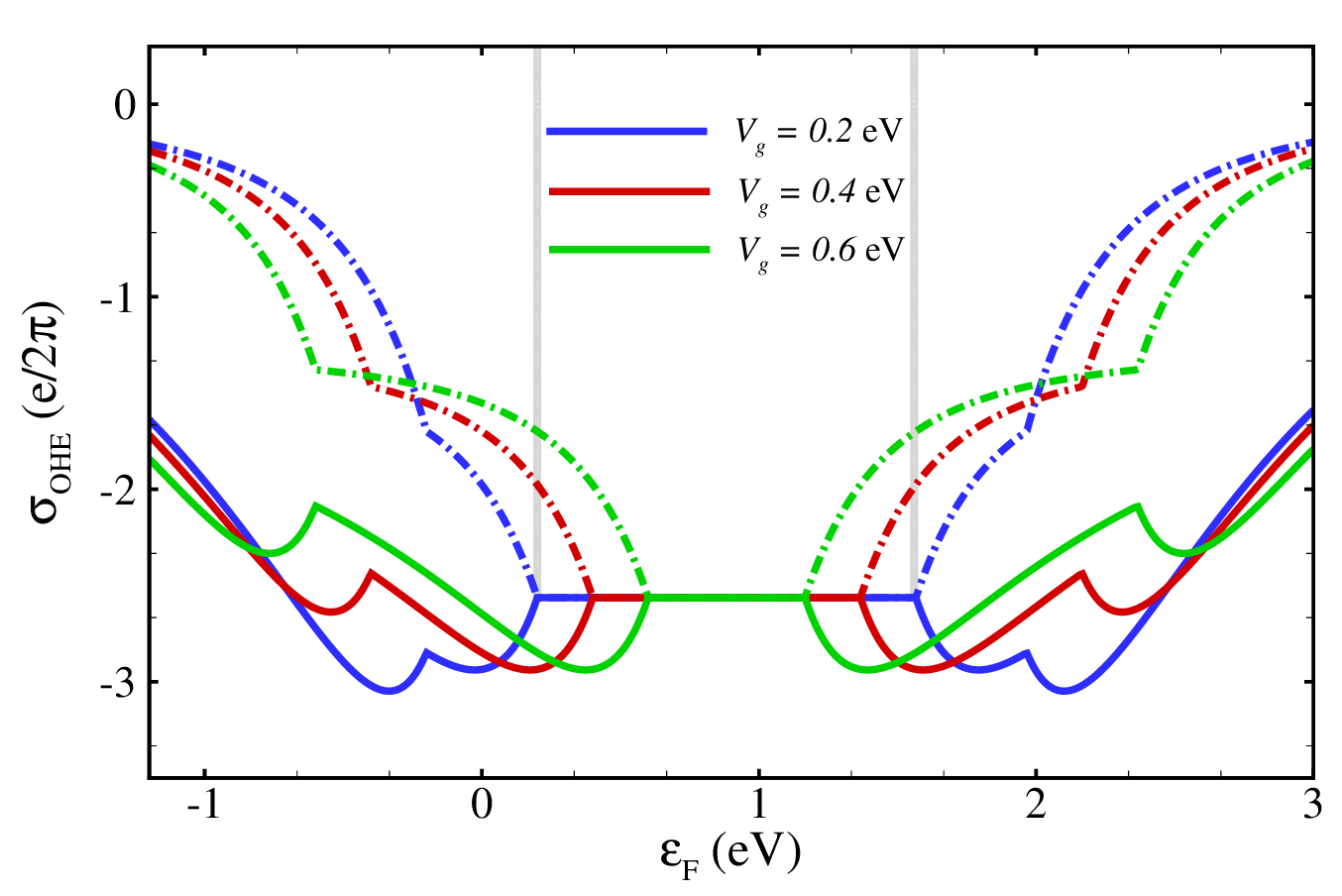}%
  }
\caption{\label{fig3} (Color online) The total orbital Hall conductivity (solid lines) and the intrinsic orbital Hall conductivity (dashed-dotted lines)  $\sigma_{\rm{OHE}}$ versus the Fermi energy $\varepsilon_{\rm F}$ for biased bilayer of 2H-MoS$_2$ for  $V_g=0.2, 0.4, $ and $0.6$ eV. The vertical grey lines delimit the band gap for $V_g=0.2$ eV. For higher Fermi energies, the extrinsic term dominates the orbital Hall conductivity for the system with broken inversion symmetry.  The overall conductivity rises to a maximum value and then progressively falls until the second band begins to fill and the same process is repeated, while the intrinsic term reduces as we move further from the gap.
    }
\end{figure}
From Eqs.~\eqref{S} and~\eqref{D}, the intrinsic contribution to orbital Hall conductivity is found as
\begin{equation}\label{int}
\sigma_{\rm{OHE}}^{\rm{int}}=-\sum_i\frac{g_sg_v}{6\hbar^2}\frac{em_e}{2\pi}\frac{m^2(\hbar v_{\rm F})^2}{\lambda_{k_{\rm F}^i}^3}
\end{equation}
where the sum is over the bands intersecting the Fermi energy. It can be seen from this formula that the intrinsic term has a finite value for the in-gap Fermi energy. The side-jump term has two contributions one of which stems from the anomalous driving term found from off-diagonal part of the density matrix and the other one comes from the electric field correction to the collision integral. From Eqs.~\eqref{S} and~\eqref{D'} for the first and Eq.~\eqref{sj} for the second term, the side-jump contribution to the orbital Hall conductivity is given by
\begin{equation}\label{sjj}
\sigma_{\rm{OHE}}^{\rm{sj}}=-2\sum_i\frac{g_sg_v}{\hbar^2}\frac{em_e}{2\pi}\frac{m^2{k_{\rm F}^i}^2(\hbar v_{\rm F})^4}{\lambda_{k_{\rm F}^i}^3(\lambda_{k_{\rm F}^i}^2+3m^2)}
\end{equation}
Note that although the second term is related to the scattering in the system, it is not proportional to the impurity density and it is exactly equal to the first term and therefore it doubles the contribution of the anomalous extrinsic term.

Using Eq.~\eqref{sk}, the skew scattering term contribution to the orbital Hall conductivity reads as
\begin{equation}\label{skk}
\sigma_{\rm{OHE}}^{\rm{sk}}=-\sum_i\frac{g_sg_v}{2\hbar^2}\frac{em_e}{2\pi}\frac{m^2{k_{\rm F}^i}^4(\hbar v_{\rm F})^6}{\lambda_{k_{\rm F}^i}^3(\lambda_{k_{\rm F}^i}^2+3m^2)^2}
\end{equation}
Finally, the total orbital Hall conductivity is the sum of all terms given by $\sigma_{\rm{OHE}}=\sigma_{\rm{OHE}}^{\rm{int}}+\sigma_{\rm{OHE}}^{\rm{sj}}+\allowbreak\sigma_{\rm{OHE}}^{\rm{sk}}$. Note that the results are the same as that of the monlayer MoS$_2$ (see Appendix~\ref{ap-three})with a sum over the bands intersecting the Fermi energy. In Fig.~\ref{fig3} the total orbital Hall conductivity is compared with the intrinsic orbital Hall conductivity for different values of the gate voltage $V_g$. We can see that for the inversion symmetry broken system the extra extrinsic terms give rise to the orbital Hall conductivity of the system such that it immediately becomes the main contribution to the total orbital Hall conductivity. On the other hand, close to the band gap, the extrinsic terms completely change the trend of the orbital Hall conductivity. While the intrinsic term decreases with the Fermi energy, the total conductivity increases to a maximum value and then decreases gradually until the second band starts to populate and the same  process will be repeated. 
    \begin{figure}
 \centering
\subfloat{%
  \includegraphics[width=4.in]{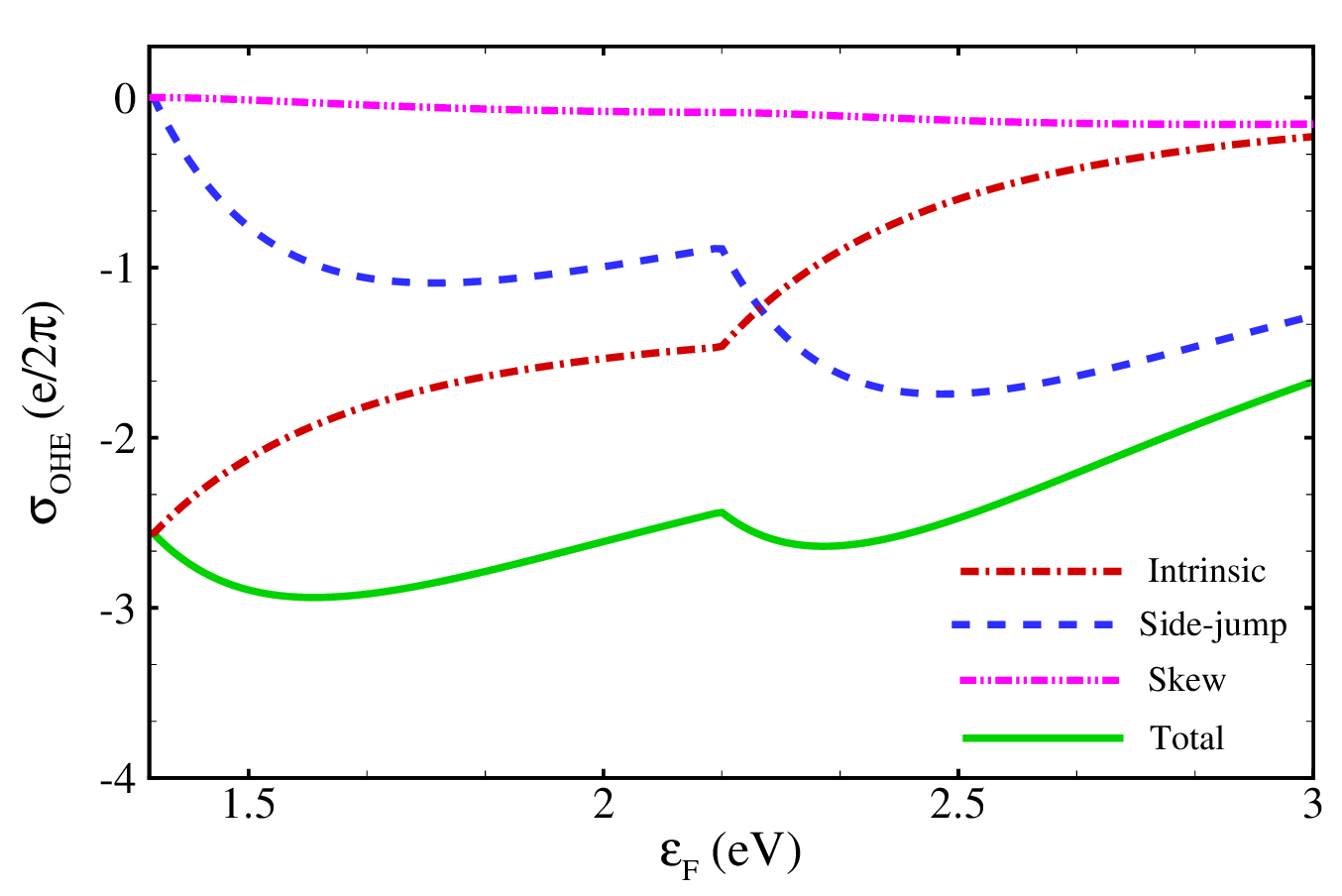}%
  }

\caption{(Color online) Intrinsic, side-jump, skew and the total orbital Hall conductivity $\sigma_{\rm{OHE}}$ versus the Fermi energy $\varepsilon_{\rm F}$ for biased bilayer of 2H-MoS$_2$ for $V_g=0.4$ eV.  }
    \label{fig4}
    \end{figure}

The individual terms of the orbital Hall conductivity for an electron-doped system are also illustrated in Fig.~\ref{fig4} for $V_g=0.4$ eV. Referring to Eqs.~\eqref{int}-\eqref{skk}, we conclude that for each band we can write $\sigma_{\rm{OHE}}^{\rm{sj}}=\frac{12}{\alpha_{k_{\rm F}}}\sigma_{\rm{OHE}}^{\rm{int}}$ and  $\sigma_{\rm{OHE}}^{\rm{sk}}=\frac{3}{\alpha_{k_{\rm F}}^2}\sigma_{\rm{OHE}}^{\rm{int}}$ with $\alpha_{k}=1+(\frac{2m}{\hbar v_{\rm F}k})^2$. It is evident from these equations that for the systems with small band gaps (other than the present system) where we can have $\hbar v_{\rm F}k_{\rm F}\gg2m$, the side-jump orbital Hall conductivity for each band can be as large as $12\sigma_{\rm{OHE}}^{\rm{int}}$ while the skew term can reach $3\sigma_{\rm{OHE}}^{\rm{int}}$. 

\section{Summary and discussions}\label{five}
In this work, we investigated orbital Hall transport in bilayer MoS$_2$ using quantum kinetic theory. While previous studies in this system have focused on the intrinsic OHE, this work introduces the extrinsic contributions in two distinct scenarios: the unbiased system with preserved inversion symmetry and the biased system where inversion symmetry is broken. We also explore how inversion symmetry breaking influences the total OHE.

Three distinct extrinsic mechanisms are identified. The first is a Fermi surface contribution to the side-jump mechanism, which arises from an anomalous driving term and contributes to the zeroth-order off-diagonal part of the density matrix. Additionally, corrections to the band-diagonal density matrix at zeroth order lead to a second side-jump contribution, originating from the electric field correction to the collision integral. Lastly, we identify a skew-scattering term as another significant extrinsic contribution.
  
\begin{figure}[h]
\centering
\subfloat{%
  \includegraphics[width=4.in]{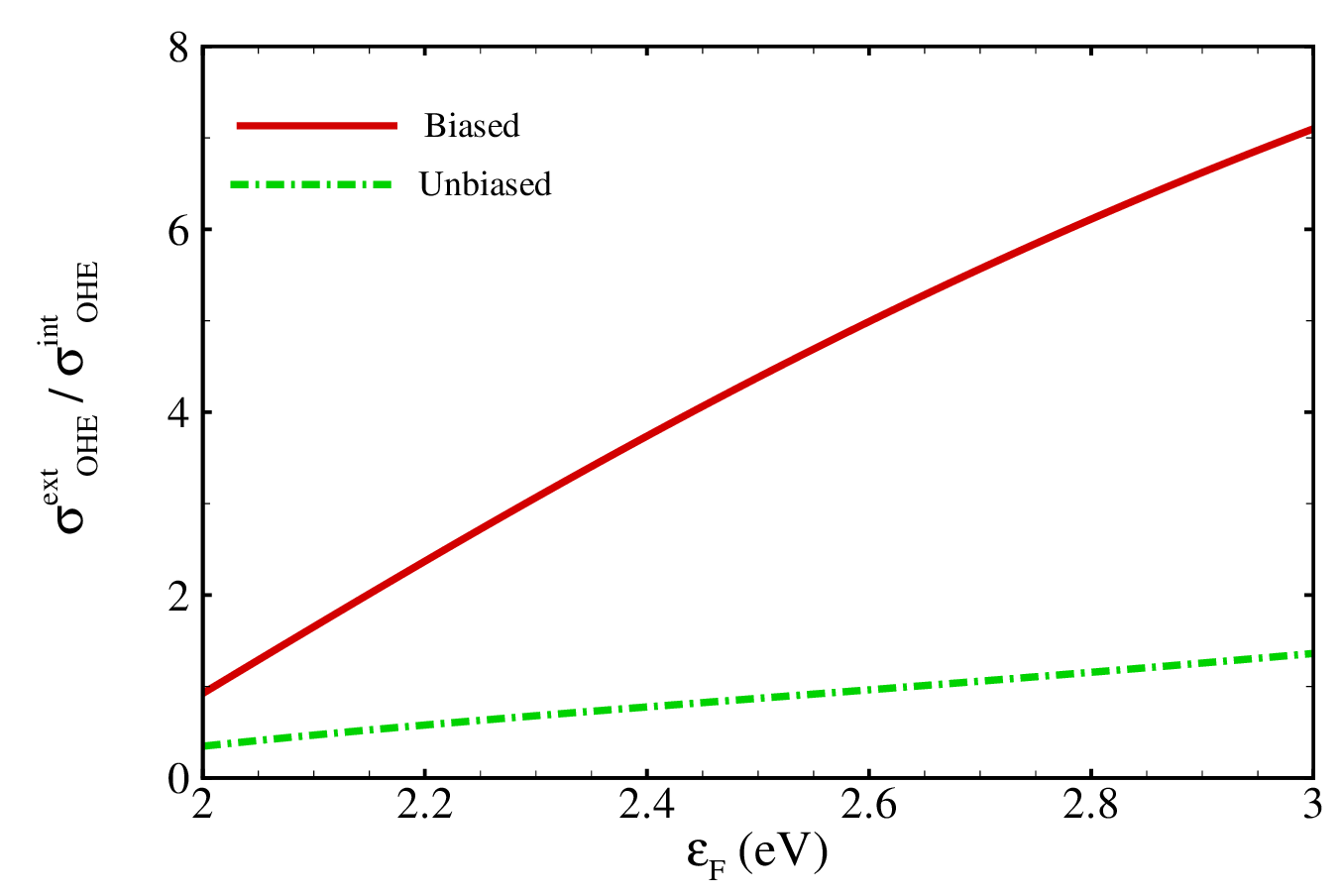}%
  }
\caption{\label{fig5} (Color online) $\sigma_{\rm{OHE}}^{\rm{ext}}/\sigma_{\rm{OHE}}^{\rm{int}}$ versus the Fermi energy $\varepsilon_{\rm F}$ for unbiased and biased ($V_g=0.2$ eV) 2H-MoS$_2$ bilayer in electron-doped region. Notice that because the diagonal elements of the OAM or intraband terms disappear in the centrosymmetric system, the absolute value of $\sigma_{\rm{OHE}}^{\rm{ext}}/\sigma_{\rm{OHE}}^{\rm{int}}$ is lower than in the biased noncentrosymmetric bilayer.
    }
\end{figure}

We find that although the extrinsic term significantly affects the orbital Hall conductivity $\sigma_{\rm{OHE}}$ of the inversion-symmetric bilayer, the absolute value of $\sigma_{\rm{OHE}}^{\rm{ext}}/\sigma_{\rm{OHE}}^{\rm{int}}$ is smaller compared to the biased noncentrosymmetric bilayer. This arises because in the centrosymmetric system, the diagonal elements of the OAM or intraband terms vanish, as expected. Consequently, the non-Abelian structure of the OAM results in an orbital Hall conductivity operator with vanishing diagonal elements. Therefore, the extrinsic orbital Hall conductivity is limited to contributions from the interband terms of the density matrix, specifically the anomalous extrinsic term.  
 This results in domination of the intrinsic part for smaller but still considerable Fermi energies. But even in this scenario, the extrinsic contribution remains significant for larger Fermi energies. For instance, in the electron-doped case, the extrinsic term begins to exceed the intrinsic contribution at $\varepsilon_{\rm F}\approx 2.64$ eV, eventually reaching about 1.4 times the intrinsic value for $\varepsilon_{\rm F}=3$ eV (see Fig.~\ref{fig5}). A carrier density of $10^{13}-10^{14}$ cm$^{-2}$ is needed to span this Fermi energy range in realistic systems which can be achieved employing ionic liquid gating~\cite{brumme2015first}. Interestingly, in the hole-doped case, the extrinsic contribution changes sign with increasing Fermi energy. Since this contribution eventually becomes larger than the intrinsic term, the total orbital Hall conductivity decreases strongly and can even experience a sign change upon increasing the enery. 
 When the inversion symmetry is broken by a gate voltage, the interlayer hopping does not affect the eigenvectors and eigenvalues to first-order perturbation theory. As a result, the layers' Hilbert spaces decouple, and the intraband elements of the OAM acquire finite values. For the special case of massive Dirac-like Hamiltonians, the off-diagonal terms vanish (Eq.~\eqref{mm}). This allows both the diagonal and off-diagonal parts of the density matrix to play significant roles in the extrinsic orbital Hall conductivity, leading to a substantial enhancement such that in this case, for a small gate voltage of 0.2 eV, the extrinsic contribution can exceed the intrinsic term by more than 7 times at $\varepsilon_{\rm F}=3$ eV (see Fig.~\ref{fig5}).
  
Notably, this difference in extrinsic contributions between the unbiased and biased cases arises from the fundamentally distinct nature of the OMM and OAM operators. For smaller values of the interlayer hopping, even a small gate voltage can drastically alter the value and behavior of the total orbital Hall conductivity. It is also important to note that, as all extrinsic terms are Fermi surface effects, the in-gap orbital Hall plateau remains robust against extrinsic contributions.

Our calculations highlight the significant and less-studied role of disorder-induced terms in the OHE of bilayer MoS$_2$. We demonstrate that while the extrinsic contribution surpasses the intrinsic one in both centrosymmetric and noncentrosymmetric systems, breaking inversion symmetry significantly enhances the extrinsic orbital Hall conductivity. These findings could be extended to other TMD bilayers with different band structures.

In this work, we assumed short-range disorder; a potential extension could involve considering long-range Coulomb potentials. Given that impurities are almost always present in real systems, we propose that our approach for tuning the orbital Hall conductivity could inspire experimental efforts to characterize the OHE in bilayer TMDs and explore its application in orbitronics. It is important to highlight that Coulomb potentials exhibit greater smoothness compared to short-range potentials, which diminishes the asymmetry observed in scattering events. This smooth characteristic mitigates the intensity of skew scattering, as it depends on pronounced potential gradients to induce asymmetry, thereby suggesting a decrease in the overall contribution to the net transverse current. Additionally, Coulomb disorder results in a more uniform potential landscape, which minimizes abrupt alterations in the trajectory of carriers. Consequently, this could lead to a reduction in the transverse displacement linked to the side-jump effect. As a result, Coulomb impurity scattering might tend to reduce the contributions of side-jump and skew scattering to the orbital Hall conductivity.

Although extrinsic effects dominate in many cases, intrinsic mechanisms can still play a substantial role in specific systems. The enhanced OHE observed in noncentrosymmetric materials has profound implications for both fundamental research and practical applications, suggesting these materials may be more suitable for directly observing the OHE, which remains elusive in experiments despite strong theoretical predictions.

Further research is necessary to comprehensively understand the relative contributions of intrinsic and extrinsic effects across various materials and conditions. Such insights could pave the way for novel spin-orbitronic devices, advancing the field of orbitronics and leveraging the unique properties of TMD bilayers.

\section*{Acknowledgements}
This work is based upon research funded by Iran National Science Foundation (INSF) under project No.4024953.

\appendix

\section{Orbital Hall current operator of centrosymmetric bilayer TMD in perturbative scheme}\label{ap-one}
Considering the hopping terms as a perturbation to the decoupled Hamiltonian of the bilayer, from the degenerate perturbation theory, the eigenvectors of the system are found as
\begin{eqnarray}
&&\ket{\phi_{-,v,\tau}}=\frac{1}{2}\Bigl(-\tau\xi_k^{-}e^{-i\tau\varphi},\xi_k^{+},\tau\xi_k^{-}e^{i\tau\varphi},-\xi_k^{+} \Bigr)^T,\label{ev1}\\
&&\ket{\phi_{+,v,\tau}}=\frac{1}{2}\Bigl(-\tau\xi_k^{-}e^{-i\tau\varphi},\xi_k^{+},-\tau\xi_k^{-}e^{i\tau\varphi},\xi_k^{+} \Bigr)^T,\label{ev2}
\end{eqnarray}
for the valence bands ($v$) and for the conduction bands ($c$) we have
\begin{eqnarray}
&&\ket{\phi_{-,c,\tau}}=\frac{1}{2}\Bigl(\tau\xi_k^{+}e^{-i\tau\varphi},\xi_k^{-},-\tau\xi_k^{+}e^{i\tau\varphi},-\xi_k^{-}\Bigr)^T,\\\label{ev3}
&&\ket{\phi_{+,c,\tau}}=\frac{1}{2}\Bigl(\tau\xi_k^{+}e^{-i\tau\varphi},\xi_k^{-},\tau\xi_k^{+}e^{i\tau\varphi},\xi_k^{-} \Bigr)^T,\label{ev4}
\end{eqnarray}
where $\xi_k^{\pm}=\sqrt{1\pm\xi_k}$ and $\xi_k=m/\lambda_k$ with $\lambda_k=\sqrt{(\hbar v_{\rm F}k)^2+m^2}$ and $\varphi$ being the angle between $\bm k$ and $\hat{x}$. Here $+$($-$) subscripts in $\ket{\phi_{+(-),c,\tau}}$ indicates the bonding (antibonding) linear combination of the unperturbed eigenvectors of the two layers on valence and conduction band subspace such that $|\phi_{\pm,c\left(v\right),\tau}\rangle=1/\sqrt2(|\psi_{2,c(v),\tau}\rangle\pm|\psi_{1,c(v),\tau}\rangle)$ where $|\psi_{1(2),c(v),\tau}\rangle$ are the eigenvectors of the unperturbed Hamiltonian~\cite{cysne2022orbital}.

Considering the first order corrections in interlayer coupling $t_{\bot}$, the eigenvalues of the valence and conduction bands are also found as
\begin{eqnarray}
&&\varepsilon_{v,\pm}(k)=m-\lambda_k\pm\frac{t_{\bot}}{2}(1+\xi_k),\\
&&\varepsilon_{c,\pm}(k)=m+\lambda_k\pm\frac{t_{\bot}}{2}(1-\xi_k).
\end{eqnarray} 
In this case, on the basis of Bloch bands or the eigenvectors of the unbiased system~\eqref{ev1}-\eqref{ev4}, the OMM operator is given by
\begin{equation}
\hat{{\bm M}}_z(k)=\begin{pmatrix}
\hat{M}^v_z(k)&\hat{0}_{2\times2}\\ \hat{0}_{2\times2}&\hat{M}^c_z(k)
\end{pmatrix},
\end{equation}
where $\hat{0}_{2\times2}$ is the $2\times2$ zero matrix and $\hat{M}^v_z(k)$ and $\hat{M}^c_z(k)$ are calculated in the subspace of valence and conduction respectively and are given by $\hat{M}^v_z(k)=\hat{M}^c_z(k)=\tau M_0(k)\begin{pmatrix}
0&1\\ 1&0\end{pmatrix}$ with $M_0(k)=\frac{e}{2\hbar}\frac{(\hbar v_{\rm F})^2m}{((\hbar v_{\rm F}k)^2+m^2)}$ being the OMM in massive Dirac model. 

The orbital Hall current operator $\hat{\bm J}_y$ can be found from
\begin{equation}
\begin{split}
&\hat{{\bm J}}_y=\tau \beta\left(k\right)\times\\
&\begin{pmatrix}
0&-\xi_k^+\xi_k^-\sin{\varphi}&i\cos{\varphi}&\xi_k\sin{\varphi}\\ 
-\xi_k^+\xi_k^-\sin{\varphi}&0&\xi_k\sin{\varphi}&i\cos{\varphi}\\ 
i\cos{\varphi}&\xi_k\sin{\varphi}&0&\xi_k^+\xi_k^-\sin{\varphi}\\
\xi_k\sin{\varphi}&i\cos{\varphi}&\xi_k^+\xi_k^-\sin{\varphi}&0
\end{pmatrix},
\end{split}
\end{equation}
with $\beta\left(k\right)=-\frac{m_e v_{\rm F}^2}{k}\xi_k\xi_k^+\xi_k^-$.
\section{Lower boundary of gate voltage for validity of perturbative scheme in biased bilayer TMD}\label{ap-two}
\begin{figure}[h]
 \captionsetup[subfigure]{labelformat=empty}
\centering
  \subfloat[]{\includegraphics[width=0.4\linewidth]{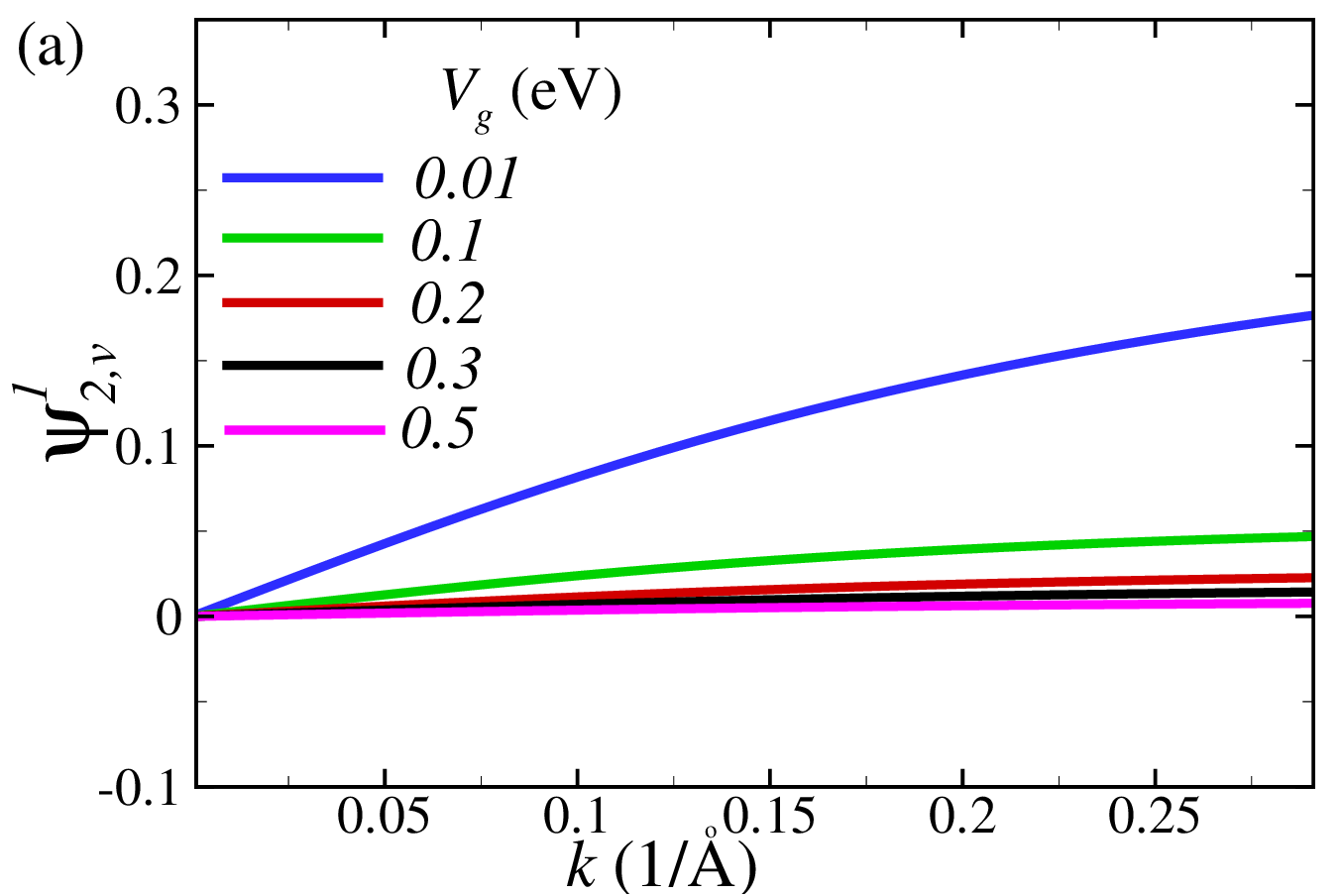}}
\hfil
    \subfloat[]{\includegraphics[width=0.4\linewidth]{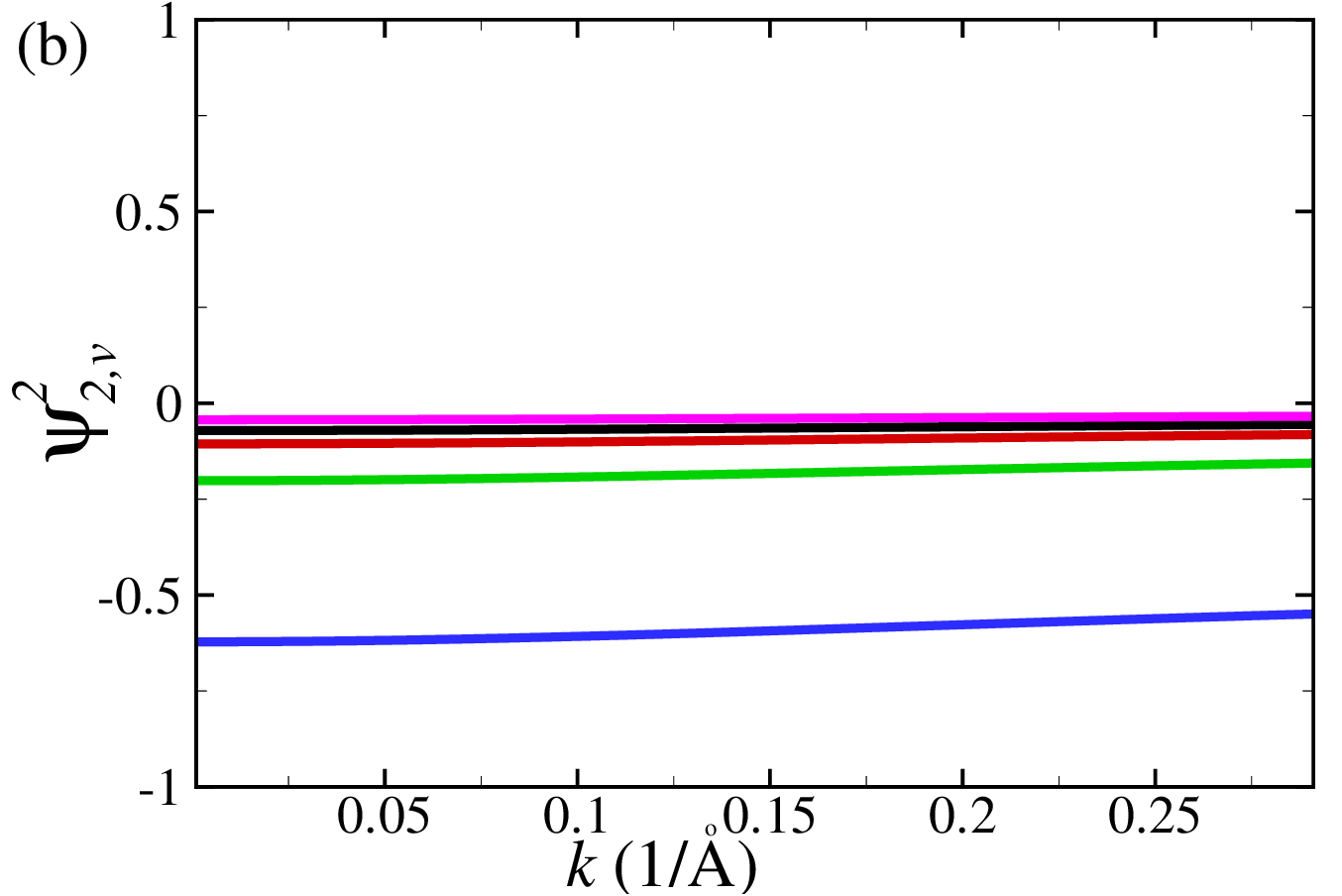}}
    \par
\negthickspace\negthickspace\negthickspace\negthickspace\negthickspace\negthickspace
  
    \subfloat[]{\includegraphics[width=.4\linewidth]{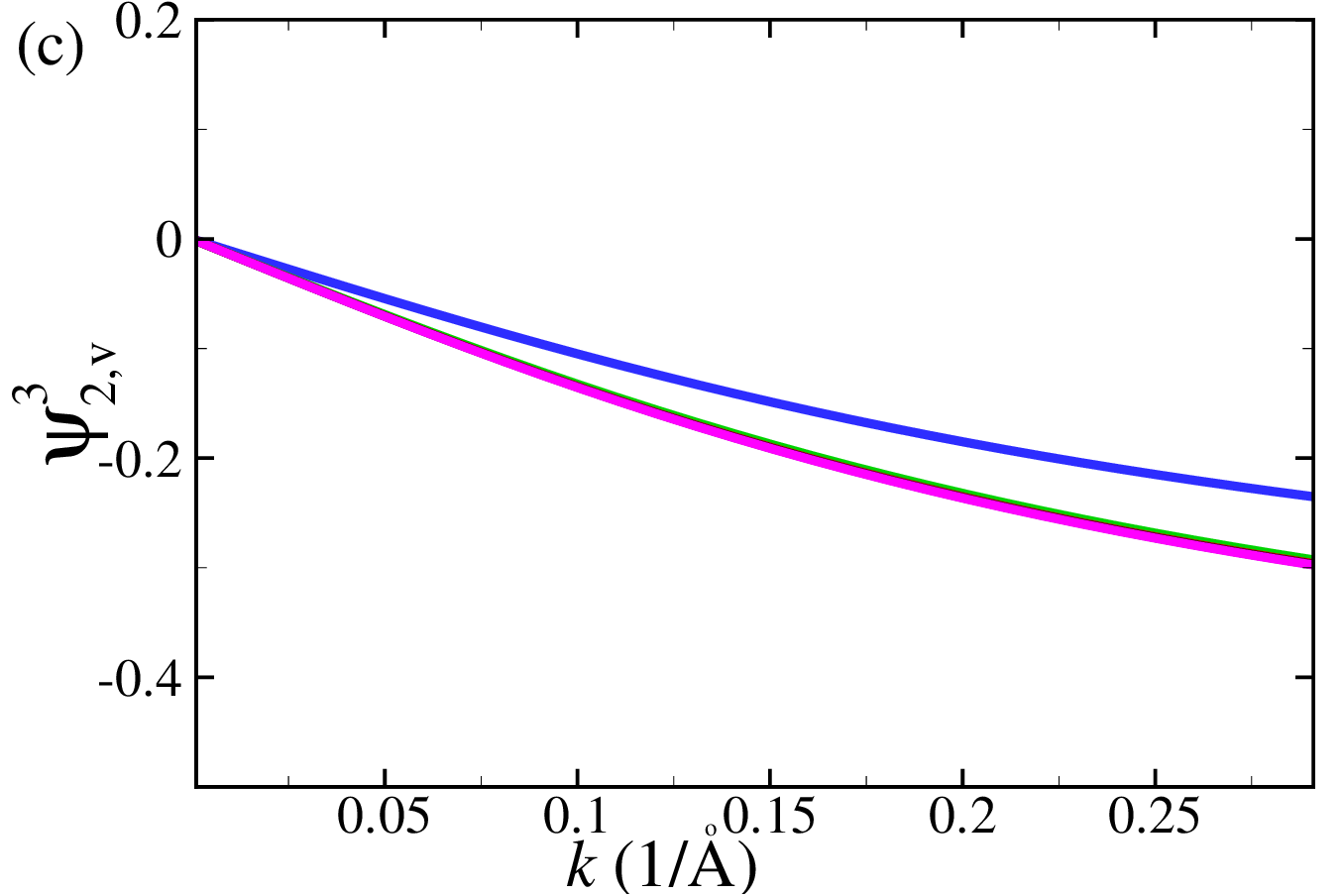}}
\hfil
    \subfloat[]{\includegraphics[width=0.4\linewidth]{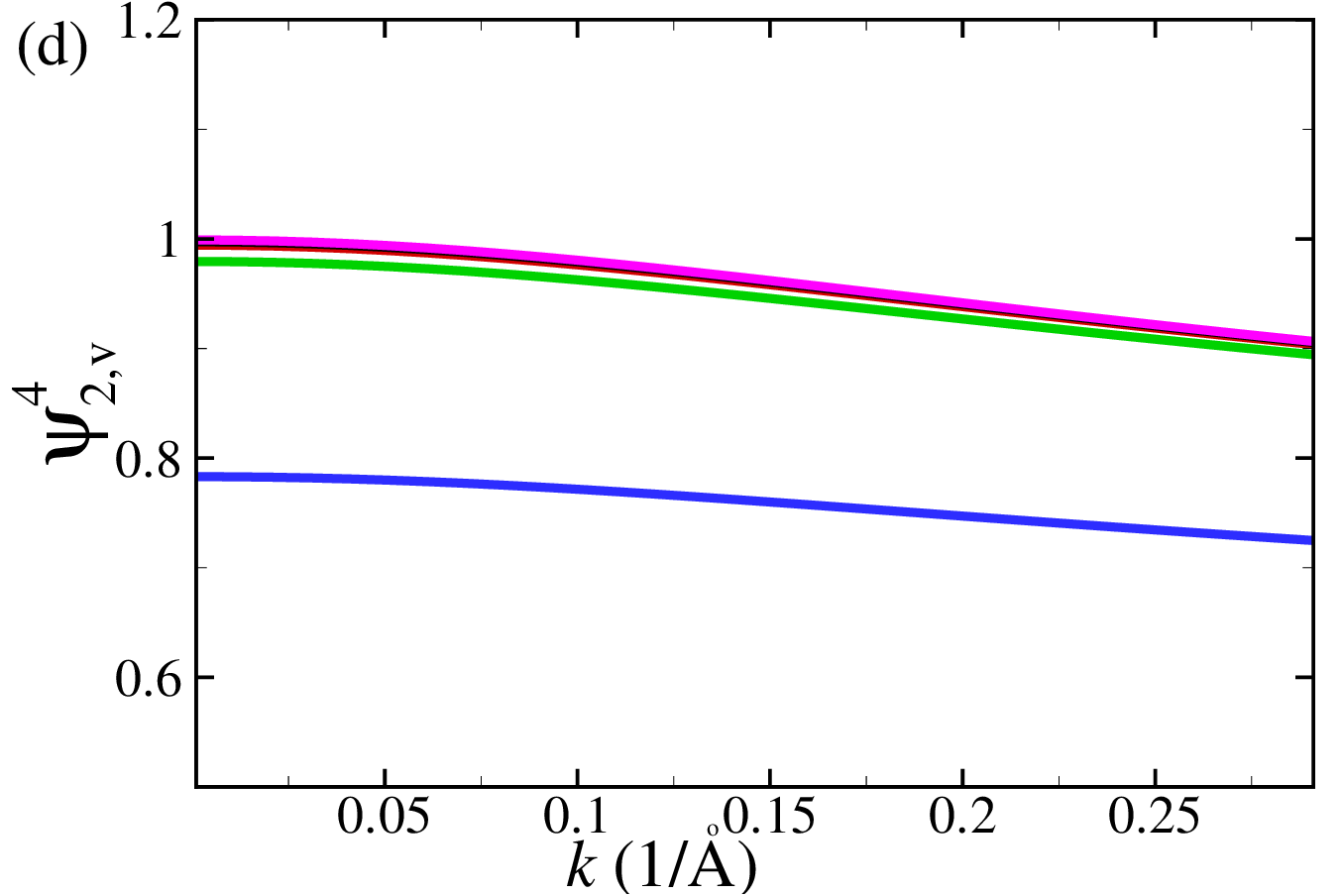}}

\caption{(Color online) (a)-(d) The four components of the exact eigenvector $\ket{\psi_{2,v,\tau}}$ obtained by numerical diagonalization of Hamiltonian of Eq.~\eqref{HU} as a function of the wave vector $k$ for $V_g=0.01, 0.1, 0.2, 0.3$ and $0.5$ eV. \label{fig6}    }
 \end{figure}

Here we have plotted the four components of the exact eigenvector $\ket{\psi_{2,v,\tau}}$ by diagonalizing the Hamiltonian of Eq.~\eqref{HU} for different values of the gate voltage. We can see that for $V_g\geqslant0.2$ eV the components are approximately unchanged when the gate voltage varies. As $V_g$ dose not appear in the eigenvectors found from the perturbative scheme (Eqs.~\eqref{uev1}-\eqref{uev4}), we conclude that for this range of gate voltages, the perturbative treatment used in the text works properly. The same argument holds for other eigenvectors as well (See Fig.~\ref{fig6}).

\section{OHE in monolayer MoS$_2$}\label{ap-three}
For a monolayer MoS$_2$, the Hamiltonian is given by
\begin{equation}
{\cal H}\left(k\right)=\hbar v_{\rm F}(\tau k_x\sigma_x-k_y\sigma_y)+m\sigma_z
\end{equation}
where $\tau=\pm$ is the valley index, $s_z$ is the Pauli matrix in the spin space, $\gamma_{\pm}=\hbar v_{\rm F}(\tau k_x\pm ik_y)$ and $\hbar v_{\rm F}=at$ with $a=3.160\,\rm{\AA}$ the lattice constant, $t=1.137\,\rm{eV}$ the nearest-neighbor  hopping and $2m=1.766\,\rm{eV}$ the band gap. The eigenvalues are given by $\varepsilon_k=\pm\sqrt{(\hbar v_{\rm F}k)^2+m^2}$ where $\pm$ correspond to the conduction and valence bands. The eigenvalues are also given by
\begin{equation}\label{cv}
\ket{\psi_{\pm,\tau}}=\frac{1}{\sqrt{2}}\Bigl(\pm\tau\xi_k^{\pm}e^{i\tau\varphi},\xi_k^{\mp}\Bigr)^T,
\end{equation}
 with $\xi_k^{\pm}=\sqrt{1\pm\xi_k}$ and $\xi_k=m/\lambda_k$ with $\lambda_k=\sqrt{(\hbar v_{\rm F}k)^2+m^2}$ as before.
\begin{figure}[h]
\centering
\subfloat{%
  \includegraphics[width=4.in]{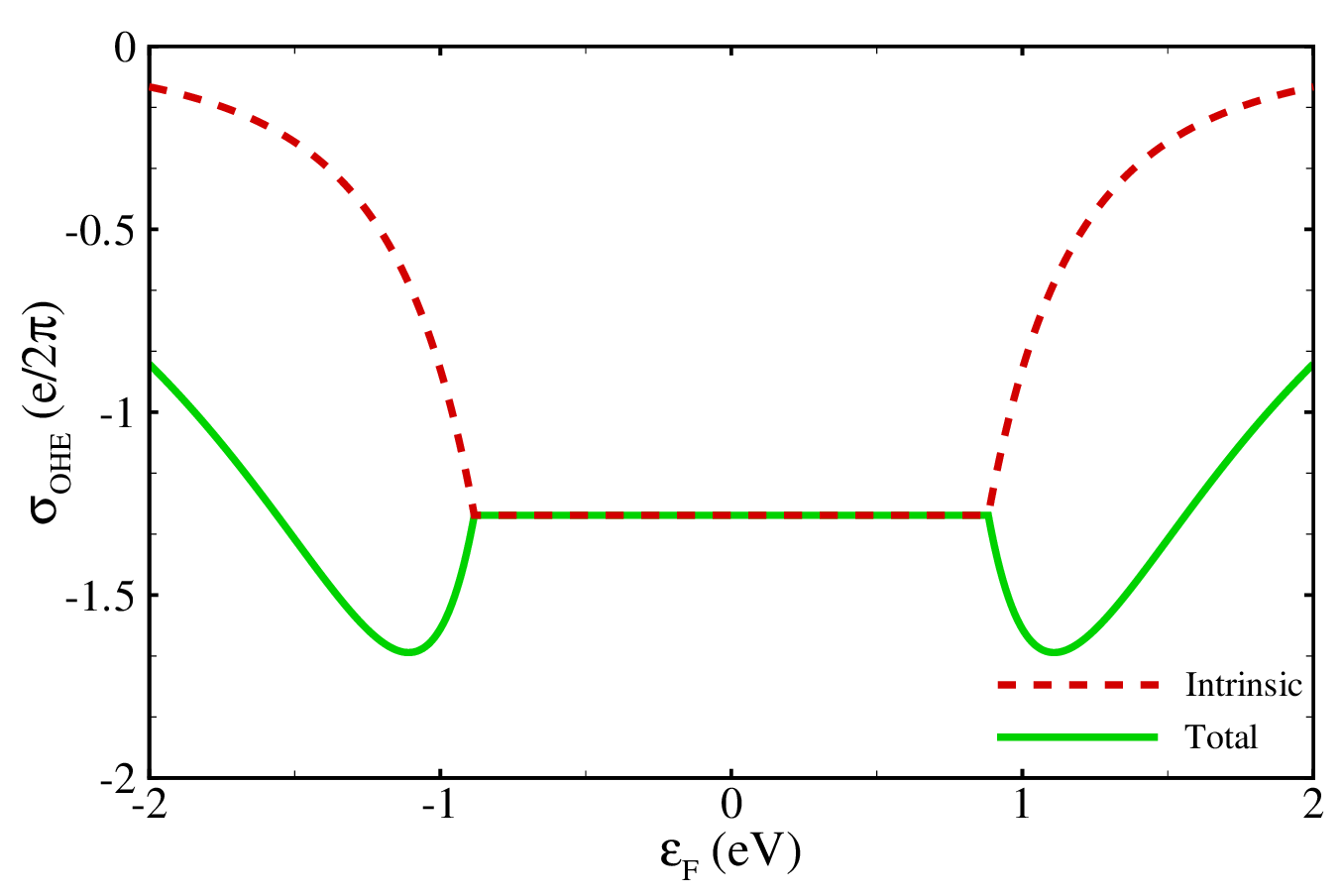}%
  }
\caption{\label{fig7} (Color online) The intrinsic (dashed red line) and the total orbital Hall conductivity (solid green line)   $\sigma_{\rm{OHE}}$ versus the Fermi energy $\varepsilon_{\rm F}$ for monolayer MoS$_2$.
    }
\end{figure}
 
Taking the same steps as in bilayer system, for a monolayer MoS$_2$, the intrinsic contribution to the orbital Hall conductivity is given by
 \begin{equation}\label{int1}
\sigma_{\rm{OHE}}^{\rm{int}}=-\frac{g_sg_v}{6\hbar^2}\frac{em_e}{2\pi}\frac{m^2(\hbar v_{\rm F})^2}{\lambda_{k_{\rm F}}^3}.
\end{equation}
For the side-jump contribution we have
\begin{equation}\label{sjj1}
\sigma_{\rm{OHE}}^{\rm{sj}}=-2\frac{g_sg_v}{\hbar^2}\frac{em_e}{2\pi}\frac{m^2{k_{\rm F}}^2(\hbar v_{\rm F})^4}{\lambda_{k_{\rm F}}^3(\lambda_{k_{\rm F}}^2+3m^2)}.
\end{equation}
The skew scattering term reads as
\begin{equation}\label{skk1}
\sigma_{\rm{OHE}}^{\rm{sk}}=-\frac{g_sg_v}{2\hbar^2}\frac{em_e}{2\pi}\frac{m^2{k_{\rm F}}^4(\hbar v_{\rm F})^6}{\lambda_{k_{\rm F}}^3(\lambda_{k_{\rm F}}^2+3m^2)^2}.
\end{equation}
The intrinsic and the total orbital Hall conductivity are shown in Fig.~\ref{fig7}.

Similar to the biased bilayer, here also we find that  $\sigma_{\rm{OHE}}^{\rm{sj}}=\frac{12}{\alpha_{k_{\rm F}}}\sigma_{\rm{OHE}}^{\rm{int}}$ and  $\sigma_{\rm{OHE}}^{\rm{sk}}=\frac{3}{\alpha_{k_{\rm F}}^2}\sigma_{\rm{OHE}}^{\rm{int}}$ with $\alpha_{k}=1+(\frac{2m}{\hbar v_{\rm F}k})^2$. For very large Fermi energies where $\hbar v_{\rm F}k_{\rm F}\gg2m$, the side-jump orbital Hall conductivity for each band can be as large as $12\sigma_{\rm{OHE}}^{\rm{int}}$ while the skew term can reach $3\sigma_{\rm{OHE}}^{\rm{int}}$ (although for monolayer TMDs, the model Hamiltonian may not work properly in this regime). Our results differ somewhat from those reported in Ref.~\cite{liu2024dominance} such that we have an extra $\alpha_{k_{\rm F}}$ in the denominator of the side-jump term and also an extra $3\alpha_{k_{\rm F}}^2$ in the denominator of the skew term. Apart from the factor $1/3$ in skew term, it seems that in Ref.~\cite{liu2024dominance} ,the calculations have been performed for $m\rightarrow0$ which is obviously not the case in monolayer MoS$_2$. 

\bibliographystyle{elsarticle-num} 
\bibliography{ref}






\end{document}